\documentclass[12pt]{article}

\usepackage{graphicx}
\usepackage{dcolumn}
\usepackage{bm}
\usepackage[a4paper,top=3.5cm,bottom=3.2cm,left=2.3cm,right=2.3cm,bindingoffset=5mm]{geometry}
\usepackage{amsfonts}
\usepackage{latexsym}
\usepackage{amsmath}
\usepackage{amssymb}
\usepackage[english]{babel}
\usepackage[utf8]{inputenc}
\usepackage[T1]{fontenc}
\usepackage{slashed}
\usepackage{float}
\usepackage{subfigure}
\usepackage{color}
\usepackage{multirow}
\usepackage{cite}

\DeclareUnicodeCharacter{2212}{-}

\begin{document}

\vspace*{-15mm}
\begin{flushright}
SISSA 09/2022/FISI\\
IPMU22-0015
\end{flushright}
\vspace*{5mm}

\vspace{1.0cm}
\begin{center}
{\bf{\large Measurements of $\mu\to 3e$ Decay with Polarised Muons 
as a Probe of New Physics
}}\\

\vspace{0.4cm}
P. D. Bolton$\mbox{}^{a),}$\footnote{Email: \texttt{patrick.bolton@ts.infn.it}} and S. T. Petcov$\mbox{}^{a,b),}$\footnote{Also at:
Institute of Nuclear Research and Nuclear Energy,
Bulgarian Academy of Sciences, 1784 Sofia, Bulgaria.} 
\\[1mm]
\end{center}
\vspace*{0.50cm}
\centerline{$^{a}$ \it SISSA/INFN, Via Bonomea 265, 34136 Trieste, Italy}
\vspace*{0.2cm}
\centerline{$^{b}$ \it Kavli IPMU (WPI), UTIAS,
The University of Tokyo, Kashiwa, Chiba 277-8583, Japan}

\vspace*{0.8cm}

\begin{abstract}
Working within the Standard Model  
effective field theory approach, we examine the possibility to test charged lepton flavour violating (cLFV) new physics (NP) by angular measurements of the outgoing electron and positrons in the polarised $\mu^{+}\to e^{+}e^{+}e^{-}$ decay. This decay is planned to be studied with a sensitivity to the branching ratio of 
$B_{\mu\to3e}\sim 10^{-15}$ by the Mu3e experiment at PSI. To illustrate the potential of these measurements,
we consider a set of eight effective 
operators generating the decay at the tree level, including dimension-five dipole operators and dimension-six scalar and vector four-fermion operators of different chiralities.
We show that if the polarised 
$\mu^{+}\to e^{+}e^{+}e^{-}$ decay is observed and is induced by a single operator, data from three angular observables -- two P-odd asymmetries and one T-odd asymmetry -- 
can allow to discriminate between all considered 
operators with the exception of scalar and 
vector operators of opposite chirality. For these two types of operators, the P-odd asymmetry 
is maximal in magnitude; they can be distinguished, in principle, by measuring the helicities of the outgoing positrons. The observation of a non-zero T-odd asymmetry would indicate the presence in the decay rate of a dipole and vector operator
interference term that is T (CP) violating.
\end{abstract}

\newpage

\section{Introduction}
\label{sec:int}

The observation of charged lepton flavour violating (cLFV) processes such as
$\mu \to e\gamma $, $\mu \to 3e$ and $\tau \to 3\mu $ decays and
$\mu - e$ conversion in nuclei would be compelling evidence for the
existence of new physics (NP) beyond the Standard Model (SM). These processes
are forbidden if the individual lepton charges $L_{\ell}$, with
$\ell =e,\mu ,\tau $, are exactly conserved. We know, however, that this is
not the case from the observation of oscillations between the different
flavour neutrinos (antineutrinos) $\nu _{\ell }\to \nu _{\ell '}$ ($
\bar{\nu}_{\ell }\to \bar{\nu}_{\ell '}$), $\ell \neq \ell '$,
$\ell ,\ell ' = e,\mu ,\tau $, triggered by non-zero neutrino masses and
mixing, which imply that $L_{\ell }\neq const$. This suggests that the
cLFV processes should take place at some level. Actually, in any theoretical
model of neutrino mass generation that correctly describes the neutrino
oscillation data, the cLFV processes should be allowed and, depending on
the specific technical aspects of the model, their rates can be predicted
to lie within certain ranges. The minimal model of this class is the SM
(renormalisable) extension, which includes three $\text{SU}(2)_{L}$ singlet
right-handed (RH) neutrino fields $\nu _{\ell R}$ whose couplings respect
the SM $\text{SU}(2)_{L}\times \text{U}(1)_{Y}$ gauge symmetry and conserve
the total lepton charge $L=L_{e} + L_{\mu }+ L_{\tau}$. In this scenario, the
neutrinos get a Dirac mass term after the spontaneous breaking of the SM
symmetry from a Yukawa-type coupling between the left-handed (LH)
lepton doublets, Higgs doublet and RH neutrinos, and the neutrinos
with definite mass $\nu _{i}$, $i=1,2,3$, are Dirac fermions. Their masses
$m_{i}$ must satisfy the constraints from existing data: the upper bound
on the $\bar{\nu}_{e}$ mass,
$m_{\bar{\nu}_{e}} \simeq m_{1,2,3} < 0.8$ eV~\cite{KATRIN:2019yun,Aker:2021gma},
and the astrophysical and cosmological upper limits on
$\sum _{i} m_{i}$~\cite{Planck:2018vyg,ParticleDataGroup:2020ssz}. The
amplitudes of cLFV processes such as $\mu \to e\gamma $,
$\mu \to 3e$ and $\tau \to 3\mu $ decays are generated at the one-loop level by diagrams
involving virtual neutrinos $\nu _{i}$~(see, e.g., Ref.~\cite{Bilenky:1987ty}).
However, the rates of these processes are strongly suppressed by the factor
$m^{4}_{i}/M^{4}_{W}$~\cite{Petcov:1976ff}, where $M_{W} \simeq 80$ GeV
is the $W^{\pm}$ boson mass, which renders them effectively unobservable.
Using the results of Ref.~\cite{Petcov:1976ff} (see also Refs.~\cite{Bilenky:1977du,Cheng:1976uq})
and the current neutrino oscillation data, one finds, for example, that
the predicted $\mu \to e \gamma $ decay branching ratio is
$B_{\mu \to e \gamma} \simeq 10^{-55}$, while the current experimental
upper bound is $B_{\mu \to e\gamma} < 4.2\times 10^{-13}$~\cite{MEG:2016leq}.

It is clear, however, that if the neutrino masses in the quoted suppression
factor are replaced by, e.g., the masses $M_{j}$ of heavy neutral leptons
(or heavy Majorana neutrinos) $N_{j}$ with values in the range of a few
GeV to 1 TeV, the rates of the cLFV processes are enhanced and may be close
to the existing upper limits (collected in Table~\ref{tab:mutoeprocesses})~\cite{Bilenky:1977du,Cheng:1976uq,Marciano:1977wx,Petcov:1977ab}.
This is precisely the case (see, e.g.,~\cite{Ilakovac:1994kj,Dinh:2012bp,Alonso:2012ji,Dinh:2013vya,Abada:2018nio})
in the low-scale version of the type~I seesaw model~\cite{Minkowski:1977sc,Yanagida:1979as,GellMann:1980vs,Glashow:1979,mohapatra:1981yp},
which technically is a straightforward modification of the simple model
of neutrino masses and mixing considered above, allowing for the non-conservation
of the total lepton charge $L$ and thus providing a natural explanation
of the smallness of the neutrino masses. Via the (low- or high-scale) leptogenesis
mechanism, the type~I seesaw scenario can also account for the observed
baryon asymmetry of the universe and establishes a link between the generation
and smallness of the neutrino masses and the origin of the baryon asymmetry. There are many alternative models of neutrino mass generation (for example,
those incorporating a Higgs triplet~\cite{Konetschny:1977bn,Magg:1980ut,Cheng:1980qt},
RH fermion triplets~\cite{Foot:1988aq,Abada:2008ea}, radiative neutrino
mass generation~\cite{Zee:1980ai,Petcov:1982en,Babu:1988ki,Ma:2006km,Cai:2017jrq}
or minimal flavour violation~\cite{Cirigliano:2005ck,Davidson:2006bd,Gavela:2009cd,Alonso:2011jd,Dinh:2017smk},
etc.) in which the cLFV processes are predicted to have rates close to
the existing experimental upper limits (for a review see, e.g., Ref.~\cite{Calibbi:2017uvl}).
Being an important signal for NP, these processes have also been considered
in models with supersymmetry, grand unification and continuous and discrete
flavour symmetries~\cite{Gabbiani:1988rb,Barbieri:1995tw,Hisano:1995cp,Hisano:1998fj,Casas:2001sr,Deppisch:2004fa,Arganda:2007jw,Altarelli:2012bn,Bambhaniya:2015ipg}.

Given the number and variety of possible NP models in which the cLFV
processes are allowed and can proceed with rates close to the current
upper bounds, one can instead study the impact of NP at low energies in
the model-independent SM effective field theory. Below the scale of NP,
$\Lambda _{\text{NP}}$, but above the electroweak symmetry breaking (EWSB)
scale, one adds higher-dimensional operators to the SM Lagrangian built
from SM fields and respecting the
$\text{SU}(3)_{c} \times \text{SU}(2)_{L} \times \text{U}(1)_{Y}$ gauge symmetry,
%
\begin{align}
\mathcal{L}=\mathcal{L}_{\text{SM}}+\sum _{d\geq 5}\mathcal{L}^{(d)}\,;~~
\mathcal{L}^{(d)} = \sum _{i}
\frac{C^{(d)}_{i}}{\Lambda _{\text{NP}}^{d-4}} \mathcal{O}^{(d)}_{i}
\,,
\end{align}
where $\mathcal{L}^{(d)}$ contains effective operators
$\mathcal{O}^{(d)}_{i}$ of mass dimension $d$ with coefficients
$C^{(d)}_{i}$. An operator of dimension $d$ is suppressed by $(d-4)$ powers
of $\Lambda _{\text{NP}}$. The only operator at dimension-five is the
$|\Delta L| = 2$ Weinberg operator; additional (total) lepton number violating
operators only arise at odd higher dimensions. Lepton flavour can be broken
by operators of dimension-five and higher. At energies below the EWSB scale,
one must match the operators above onto a set of dimension $\bar{d}$ operators
constructed from degrees of freedom lighter than $M_W$ which respect the broken-phase
$\text{SU}(3)_{c} \times \text{U}(1)_{\text{em}}$ gauge symmetry. The Weinberg
and higher-dimensional $|\Delta L| = 2$ operators, for example, generate
Majorana neutrino masses, i.e.
$\mathcal{L}^{(\bar{3})}=-\frac{1}{2}M_{\nu}\bar{\nu}^{c}_{L} \nu _{L}+
\text{h.c.}$.

\begin{table}[t!]
\centering
\renewcommand{\arraystretch}{1.25}
\setlength\tabcolsep{5.2pt}
\begin{tabular}{|c|c|c|}
\hline
cLFV observable &  Current upper bound & Future sensitivity  \\  \hline\hline
$\mu^{+}\to e^{+}\gamma$ & $4.2\times 10^{-13}$~\cite{MEG:2016leq} & $6\times 10^{-14}$~\cite{MEGII:2018kmf} \\
$\mu^{+}\to e^{+}\gamma\gamma$ & $7.2\times 10^{-11}$~\cite{Bolton:1988af} & --  \\
$\mu^{+}\to e^{+}e^{+}e^{-}$ & $1.0\times 10^{-12}$~\cite{SINDRUM:1987nra} & $2.0\times 10^{-15}$~\cite{Mu3e:2020gyw} \\\hline
\multirow{4}{*}{$\mu^{-}\mathcal{N}\to e^{-}\mathcal{N}$} & $7.0\times 10^{-11}$~(S)~\cite{Badertscher:1980bt} & $\sim 10^{-17}$~(Al)~\cite{Mu2e:2014fns,COMET:2018auw} \\
 & $4.3\times 10^{-12}$~(Ti)~\cite{SINDRUMII:1993gxf} & $\sim 10^{-18}$~(Ti)~\cite{PRISM:2006abc} \\
 & $7.0\times 10^{-13}$~(Au)~\cite{SINDRUMII:2006dvw} & --\\
 & $4.6\times 10^{-11}$~(Pb)~\cite{SINDRUMII:1996fti} & --\\
\hline
\end{tabular}
\caption{Current upper bounds and future sensitivities on the branching ratios of cLFV muon decays and capture rates for muon to electron conversion in nuclei.}
\label{tab:mutoeprocesses}
\end{table}

Also generated in the low-energy effective field theory are operators which
induce cLFV processes. For example, there are 42 (48) operators with three
or four legs that trigger the $\mu \leftrightarrow e$ cLFV processes in
Table~\ref{tab:mutoeprocesses} at tree (one-loop) level~\cite{Davidson:2020hkf}.
If any of these processes are observed, the question about which operator
is generating the process will inevitably arise. The
\textit{complementarity} of $\mu \leftrightarrow e$ cLFV probes in identifying
and constraining the possible operators was investigated in Ref.~\cite{Cirigliano:2017azj,Davidson:2018kud,Davidson:2020hkf}.
Taken into account was the renormalisation group (RG) running of the operator
coefficients from the muon mass $m_{\mu}$ up to the electroweak scale,
$M_{W}$. It was found that upper bounds on $\mu \to e\gamma $,
$\mu \to e\gamma \gamma $, $\mu \to 3e$ decays and $\mu - e$ conversion in nuclei can probe orthogonal directions in the parameter space at a given
scale. A higher degree of orthogonality can help discriminate a particular
model, in which the indicated processes are triggered by specific combinations
of operators whose coefficients at $\Lambda _{\text{NP}}$ are predicted
by the model.

The quoted analyses only use upper limits on the branching ratios and
capture rates of $\mu \leftrightarrow e$ flavour changing processes. If
one or more of these processes is detected, additional information can
be obtained from the angular distributions and helicities of the final
state particles. For example, in Refs.~\cite{Okada:1999zk}, the full differential
rate for $\mu ^{+}\to e^{+}e^{+}e^{-}$ decay with a
\textit{polarised} muon was used to construct three angular distribution
asymmetries -- two P-odd and one T-odd -- sensitive to different combinations
of operators (contributing to the process at tree level). The asymmetries
were calculated in specific versions of $\text{SU}(5)$ and
$\text{SO}(10)$ SUSY grand unified theories and it was found that the sign
of the P-odd asymmetry can be different in the two models. The consideration of angular observables is particularly relevant as the
next generation of $\mu \leftrightarrow e$ cLFV search experiments are
likely to use a polarised muon beam. For example, the high-intensity muon
beams (HIMB) proposal at PSI aims to provide a flux of
$\mathcal{O}(10^{9})$ longitudinally-polarised muons per second to the
Mu3e and MEG II experiments~\cite{Aiba:2021bxe}, with an estimated average
$\mu ^{+}$ polarisation of 93\%.

In the present article, following a purely phenomenological approach, we
investigate the possibility to discriminate between different effective
operators contributing to the amplitude of the polarised
$\mu ^{+}\rightarrow e^{+}e^{+}e^{-}$ decay by using data on the three
angular observables -- two P-odd and one T-odd asymmetries. Our aim is
not to perform a comprehensive analysis of all operators that can induce
the $\mu ^{+}\rightarrow e^{+}e^{+}e^{-}$ decay, but to illustrate the
potential of the method on the example of a subset of eight commonly-used
effective operators that generate the decay at tree level. Assuming the
scale of NP $\Lambda _{\text{NP}}$ to be around 1~TeV, one can neglect to
a good approximation in the analysed P-odd and T-odd asymmetries the effects
of the renormalisation group running of the considered operators above
and below the electroweak scale. In addition to the case of having just
one operator responsible for the decay, we obtain predictions for the asymmetries
of interest also in the cases when two operators are triggering the decay.

This work is organised as follows. In Section~\ref{sec:NP}, we review
the effective operators that are commonly used to parametrise the impact
of NP above the EW scale on cLFV observables, such as the
$\mu ^{+}\to e^{+}\gamma $ and $\mu ^{+}\to e^{+}e^{+}e^{-}$ decay rates,
and that we are going to use also in our analysis. We then examine the
differential branching ratio for the $\mu ^{+}\to e^{+}e^{+}e^{-}$ decay
as a function of four kinematic variables, in particular of the two angles
defining the directions of the outgoing positrons and electron. We next
derive the total branching ratio and single differential branching ratios
in the two angles. In doing so, we define three asymmetries (two P-odd
and one T-odd) and derive their dependence on the coefficients of the effective
operators of interest. We determine further the values of these asymmetries
when only one operator, or a combination of two operators, induces
the decay. In Section~\ref{sec:disc}, we explore how a future experiment
such as Mu3e can measure these asymmetries and if so, how the NP contributing
to the cLFV process can be identified. We identify some unique signatures
and comment on how helicity measurements of the outgoing positrons and
electron can complement the angular observables. We conclude in Section~\ref{sec:conc}.

\section{New Physics Contributions to $\mu ^{+}\to e^{+}e^{+}e^{-}$ Decay Rate}
\label{sec:NP}

In the low-energy effective field theory (LEFT) of the SM, the eight operators
of lowest dimension that contribute at tree level to the cLFV
$\mu ^{+}\to e^{+}e^{+}e^{-}$ decay rate can be written, without loss of
generality (up to Fierz rearrangements), as (see, e.g.,~\cite{Kuno:1999jp,Davidson:2020hkf,Mu3e:2020gyw})
%
\begin{align}
\mathcal{L}^{(\bar{5})}+\mathcal{L}^{(\bar{6})} \supset -2\sqrt{2}G_{
\text{F}}\Big[&C_{D,L}m_{\mu}\bar{\mu} \sigma _{\alpha \beta} P_{L} e F^{
\alpha \beta} + C_{D,R} m_{\mu} \bar{\mu} \sigma _{\alpha \beta} P_{R}
e F^{\alpha \beta}
\nonumber
\\
&+C^{\ell \ell '}_{S,LL}\left (\bar{\mu} P_{L} e\right )\left (
\bar{\ell} P_{L} \ell '\right )+C^{\ell \ell '}_{S,RR}\left (
\bar{\mu} P_{R} e\right )\left (\bar{\ell} P_{R} \ell '\right )
\nonumber
\\
&+C^{\ell \ell '}_{V,LL}\left (\bar{\mu} \gamma _{\alpha} P_{L} e
\right )\left (\bar{\ell} \gamma ^{\alpha} P_{L} \ell '\right )+C^{
\ell \ell '}_{V,RR}\left (\bar{\mu}\gamma _{\alpha} P_{R} e\right )
\left (\bar{\ell}\gamma ^{\alpha} P_{R} \ell '\right )
\nonumber
\\
&+C^{\ell \ell '}_{V,LR}\left (\bar{\mu} \gamma _{\alpha} P_{L} e
\right )\left (\bar{\ell} \gamma ^{\alpha} P_{R} \ell '\right )+C^{
\ell \ell '}_{V,RL}\left (\bar{\mu} \gamma _{\alpha} P_{R} e\right )
\left (\bar{\ell} \gamma ^{\alpha} P_{L} \ell '\right )\Big]
\nonumber
\\
&+\text{h.c.}\,.
\label{eq:operators}
\end{align}
Here, the dimension-five operators with coefficients $C_{D,L}$ and
$C_{D,R}$ are dipole operators and the dimension-six are four-fermion operators.
The coefficients have been normalised to the Fermi constant
$G_{\text{F}}$; an $\mathcal{O}(1)$ coefficient therefore corresponds to
$\Lambda _{\text{NP}}\sim m_{W}$.

In the presence of the cLFV operators in Eq.~\eqref{eq:operators}, the
differential branching ratio for
$\mu ^{+}(p)\to e^{+}(k_{1})e^{+}(k_{2})e^{-}(k_{3})$ decay with an incoming
positive muon of polarisation $\varepsilon =(0,\vec{P})$, written in the
muon rest frame, can be expressed as
%
\begin{align}
\frac{dB_{\mu \to 3e}}{dx_{1} dx_{2} d\Omega _{\varepsilon}} =
\frac{3}{2\pi}\Big[&C_{1}(x_{1},x_{2})+C_{2}(x_{1},x_{2})P \cos
\theta _{\varepsilon}
\nonumber
\\
&+C_{3}(x_{1},x_{2})P \sin \theta _{\varepsilon}\cos \phi _{
\varepsilon}+C_{4}(x_{1},x_{2})P \sin \theta _{\varepsilon}\sin \phi _{
\varepsilon}\Big]\,,
\label{eq:diffB}
\end{align}
where
$d\Omega _{\varepsilon}=d\cos \theta _{\varepsilon}d\phi _{
\varepsilon}$, $P=|\vec{P}|$ is the magnitude of the muon polarisation
vector, and we define $x_{1} \equiv 2E_{k_{1}}/m_{\mu}$ and
$x_{2} \equiv 2E_{k_{2}}/m_{\mu}$, with $E_{k_{1}}$ and $E_{k_{2}}$ the
energies of the outgoing positrons in the muon rest frame (the former taken
to have larger outgoing energy, $E_{k_{1}}\geq E_{k_{2}}$). For convenience,
the $z$-axis is set to be along the outgoing electron direction
$\vec{k}_{3}$ and the decay plane is fixed to be the $x$-$z$ plane. The
angle $\theta _{\varepsilon}$ is defined to lie between $\vec{P}$ and the
outgoing electron direction, while the angle $\phi _{\varepsilon}$ is the
azimuthal rotation in the $x$-$y$ plane between $\vec{P}$ and the decay
plane. In Eq.~\eqref{eq:diffB}, the $C_{i}(x_{1},x_{2})$ are functions
of the coefficients in Eq.~\eqref{eq:operators} and are given in Appendix~\ref{app:functions}.
These expressions are obtained in the limit $m_{e} \to 0$.

The term in Eq.~\eqref{eq:dBdcosdphi} proportional to
$\cos \theta _{\varepsilon}$ is obtained from the dot product
$k_{3}\cdot \varepsilon $, while the
$\sin \theta _{\varepsilon}\cos \phi _{\varepsilon}$ term originates from
the products $k_{1}\cdot \varepsilon $ and $k_{2}\cdot \varepsilon $. On
the other hand, the term proportional to
$\sin \theta _{\varepsilon}\sin \phi _{\varepsilon}$ is generated by
$\varepsilon _{\mu \nu \rho \sigma}\varepsilon ^{\mu}k_{1}^{\nu}k_{2}^{
\rho}k_{3}^{\sigma}$, where $\varepsilon _{\mu \nu \rho \sigma}$ is the
four-dimensional Levi-Civita symbol. Under a parity (P) transformation,
the polarisation and momentum four-vectors transform as
$\varepsilon \to \varepsilon ' = (0,\vec{P})$ and
$k_{i}\to k'_{i} = (E_{k_{i}},-\vec{k}_{i})$, respectively. Under a time
(T) reversal, they instead transform as
$\varepsilon \to \varepsilon ' = (0,-\vec{P})$ and
$k_{i}\to k'_{i} = (E_{k_{i}},-\vec{k}_{i})$. As a result, the dot products
$k_{i}\cdot \varepsilon $ and the terms proportional to
$\cos \theta _{\varepsilon}$ and
$\sin \theta _{\varepsilon}\cos \phi _{\varepsilon}$ are P-odd, while
$\varepsilon _{\mu \nu \rho \sigma}\varepsilon ^{\mu}k_{1}^{\nu}k_{2}^{
\rho}k_{3}^{\sigma}$ and the term proportional to
$\sin \theta _{\varepsilon}\sin \phi _{\varepsilon}$ are T-odd. Assuming
that NP contributing to the cLFV process is invariant under CPT, the non-conservation
of T implies the violation of CP.

To obtain a double differential branching ratio in the angular variables
$\cos \theta _{\varepsilon}$ and $\phi _{\varepsilon}$, it is possible
to integrate Eq.~\eqref{eq:diffB} over the variables $x_{1}$ and
$x_{2}$. The kinematic limits of the three-body phase space in
$x_{1}$ and $x_{2}$ can be found by first examining those in the kinematic
variables $s_{1} = (p - k_{1})^{2} = m_{\mu}^{2}(1-x_{1})+m_{e}^{2}$ and
$s_{2} = (p - k_{2})^{2} = m_{\mu}^{2}(1-x_{2})+m_{e}^{2}$. These are
%
\begin{gather}
\label{eq:s2limits}
4m_{e}^{2} \leq s_{1} \leq (m_{\mu }- m_{e})^{2} \,,
\\
s_{1} \leq s_{2} \leq \frac{1}{2}\left [m^{2}_{\mu }+ 3m^{2}_{e} - s_{1}
+ m^{2}_{\mu}\,\lambda ^{\frac{1}{2}}\bigg(
\frac{m_{e}^{2}}{m_{\mu}^{2}},\frac{s_{1}}{m_{\mu}^{2}}\bigg)\lambda ^{
\frac{1}{2}}\bigg(\frac{m_{e}^{2}}{s_{1}},\frac{m_{e}^{2}}{s_{1}}
\bigg)\right ]\,,
\label{eq:s3limits}
\end{gather}
where $\lambda (x,y) = (1-x-y)^{2}-4xy$ and the condition
$s_{1} \leq s_{2}$ follows from $x_{1}\geq x_{2}$. Neglecting the electron
mass, one obtains the integration ranges $[0,1]$ and
$[1-x_{1},x_{1}]$ for $x_{1}$ and $x_{2}$, respectively. However, given
the choice $x_{2} \geq x_{1}$, one can show that the range
$[0,\frac{1}{2}]$ for $x_{1}$ is not physical, so $[\frac{1}{2},1]$ and
$[1-x_{1},x_{1}]$ can be taken as the final integration limits. These limits
lead to logarithmic divergences since the functions
$C_{i}(x_{1},x_{2})$, more specifically, the terms in
$C_{i}(x_{1},x_{2})$ proportional to the dipole coefficients
$C_{D,L}$ and $C_{D,R}$, which correspond to photon-penguin diagrams, tend
to infinity as $x_{1}\to 1$ (or equivalently, as $s_{1} \to 0 $).

In Ref.~\cite{Okada:1999zk} the collinear singularities were avoided
by introducing the cut-off parameter $\delta $ in the $x_{1}$ integration
as $[\frac{1}{2},1-\delta ]$. For all operators in Eq.~\eqref{eq:diffB}
other than the dipole operators, the $\delta \to 0$ limit can be taken
without leading to a logarithmic divergence. Non-zero $\delta $ therefore
gives small corrections at higher orders in an expansion in
$\delta $. On the other hand, the lowest order contribution for the dipole
operators is proportional to the logarithm of $\delta $. In their numerical
estimates, the authors of Ref.~\cite{Okada:1999zk} fixed
$\delta = 0.02$. To understand more exactly the dependence on
$m_{e}$, we instead perform the phase space integration for the dipole
operators over the variables $s_{1}$ and $s_{2}$, with the limits in Eqs.~\eqref{eq:s2limits}
and \eqref{eq:s3limits} and without neglecting $m_{e}$. This allowed us
to avoid introducing an undefined cut-off parameter in the calculations.
Finally, we expand the final result in the small ratio
$\hat{m}_{e} \equiv \frac{m_{e}}{m_{\mu}}$.

We obtain the double differential branching ratio as
%
\begin{align}
\label{eq:dBdcosdphi}
\frac{dB_{\mu \to 3e}}{d\Omega _{\varepsilon}} & =
\frac{B_{\mu \to 3e}}{4\pi}\Big[1+P\big(A^{\text{P}}_{\mu \to 3e}\cos
\theta _{\varepsilon}+A^{\text{P}'}_{\mu \to 3e}\sin \theta _{
\varepsilon}\cos \phi _{\varepsilon}+A^{\text{T}}_{\mu \to 3e}\sin
\theta _{\varepsilon}\sin \phi _{\varepsilon}\big)\Big]\,.
\end{align}
The total branching ratio, which can be found by integrating Eq.~\eqref{eq:dBdcosdphi}
over $\cos \theta _{\varepsilon}$ and $\phi _{\varepsilon}$ with the ranges
$[-1,1]$ and $[0,2\pi ]$, respectively, can be written as
%
\begin{align}
\label{eq:Br}
B_{\mu \to 3e} =&\,\, \frac{1}{8}|C^{ee}_{S,LL}|^{2} + 2|C^{ee}_{V,LL}|^{2}+|C^{ee}_{V,LR}|^{2}
-\frac{8}{3}|eC_{D,L}|^{2}L_{1}(\hat{m}_{e}^{2})
\nonumber
\\
&\,+ 8\,\text{Re}\big[eC_{D,L}(2C^{ee*}_{V,LL}+C^{ee*}_{V,LR})\big] + (L
\leftrightarrow R)\,,
\end{align}
where $L_{1}(x) = 17+12\ln 4x$.

Firstly, we note that the terms in the total branching ratio proportional
to $|C_{D,L}|^{2}$ and $|C_{D,R}|^{2}$ are practically equal to previous
results in the literature, cf.~Refs~\cite{Petcov:1977ab,Ilakovac:1994kj,Hisano:1998fj,Dinh:2012bp}.
Taking only the contribution from $C_{D,L(R)}$, we obtain
%
\begin{align}
\frac{B^{D,L(R)}_{\mu \to 3e}}{B^{D,L(R)}_{\mu \to e\gamma}} = -
\frac{e^{2}}{12\pi}\left (\ln (4\hat{m}_{e}^{2}\big)+\frac{17}{12}
\right ) \approx \frac{\alpha}{3\pi}\left (\ln \bigg(
\frac{m_{\mu}^{2}}{m_{e}^{2}}\bigg)-\frac{11}{4}\right )\,,
\end{align}
where we have used the result
$B_{\mu \to e\gamma} = 384\pi ^{2}(|C_{D,L}|^{2}+|C_{D,R}|^{2})$.

The differential branching ratio in $\cos \theta _{\varepsilon}$ can be
found by integrating Eq.~\eqref{eq:dBdcosdphi} over the azimuthal angle.
This gives
%
\begin{align}
\label{eq:dBdcos}
\frac{dB_{\mu \to 3e}}{d\cos \theta _{\varepsilon}} & =
\frac{B_{\mu \to 3e}}{2}\big(1+A^{\text{P}}_{\mu \to 3e}P\cos \theta _{
\varepsilon}\big)\,.
\end{align}
The P-odd asymmetry term $A^{\text{P}}_{\mu \to 3e}$ that remains is
%
\begin{align}
\label{eq:AP}
A^{\text{P}}_{\mu \to 3e} = \frac{1}{B_{\mu \to 3e}}\bigg[&\frac{1}{8}|C^{ee}_{S,LL}|^{2}
- 2|C^{ee}_{V,LL}|^{2} + \frac{1}{3}|C^{ee}_{V,LR}|^{2} - \frac{8}{3}|e
C_{D,L}|^{2} L_{2}(\hat{m}_{e}^{2})
\nonumber
\\
&- 8\,\text{Re}\big[eC_{D,L}(2C^{ee*}_{V,LL}-C^{ee*}_{V,LR})\big] - (L
\leftrightarrow R)\bigg]\,,
\end{align}
where $L_{2}(x) = 170-171\ln 2 + 12\ln 4x$. In Table~\ref{tab:AAp}, we
show the values of the asymmetry $A^{\text{P}}_{\mu \to 3e}$ when each of
the coefficients in Eq.~\eqref{eq:operators} is taken to be non-zero at
a time. The operators with coefficients $C^{ee}_{S,LL}$,
$C^{ee}_{S,RR}$, $C^{ee}_{V,LL}$ and $C^{ee}_{V,RR}$ induce maximal asymmetries,
$A^{\text{P}}_{\mu \to 3e} = \pm 1$, while those with coefficients
$C_{D,L}$, $C_{D,R}$, $C^{ee}_{V,LR}$ and $C^{ee}_{V,RL}$ have
$|A^{\text{P}}_{\mu \to 3e}| < 1$. As is evident from Eq.~\eqref{eq:AP},
the value of $A^{\text{P}}_{\mu \to 3e}$ changes sign when the replacement
$L\leftrightarrow R$ is made. We note that the only operators that interfere
at lowest order in $\hat{m}_{e}$, among those considered by us, are the
dipole and vector operators. Interference between the other operators can therefore
be neglected.

\begin{table}[t!]
\centering
\renewcommand{\arraystretch}{1.4}
\setlength\tabcolsep{5.2pt}
\begin{tabular}{|c|c|c|c|c|}
\hline
$C^{(\bar{5})}_i$/$C^{(\bar{6})}_i$ &~\cite{Okada:1999zk} &~\cite{Davidson:2020hkf} & $A^{\text{P}}_{\mu\to 3e}$ & $A^{\text{P}'}_{\mu\to 3e}$ \\  \hline\hline
$C_{D,L(R)}$ & $A_{R(L)}$ & $C_{D,R(L)}$ & $\pm 1 \pm \frac{9(17-19\ln 2)}{17+12\ln(4\hat{m}_e^2)}\approx \pm 0.63$ & $\frac{\pm 144}{5(17+12\ln(4\hat{m}_e^2))}\approx \mp 0.31$ \\
$C^{ee}_{S,LL(RR)}$ & $g_{1(2)}$ & $C^{ee}_{S,RR(LL)}$ & $\pm1$ & $0$ \\
$C^{ee}_{V,LL(RR)}$ & $g_{4(3)}$ & $C^{ee}_{V,LL(RR)}$ & $\mp 1$ & $0$ \\
$C^{ee}_{V,LR(RL)}$ & $g_{6(5)}$ & $C^{ee}_{V,LR(RL)}$ & $\pm \frac{1}{3}$ & $\pm\frac{32}{105}$  \\\hline
\end{tabular}
\caption{Values of the asymmetries $A^{\text{P}}_{\mu\to 3e}$ and $A^{\text{P}'}_{\mu\to 3e}$ when each of the coefficients of the operators in Eq.~\eqref{eq:operators} are taken to be non-zero at a time, with the equivalent coefficients of Refs.~\cite{Okada:1999zk,Davidson:2020hkf} shown for clarity.}
\label{tab:AAp}
\end{table}

One can obtain the differential branching ratio in the azimuthal angle
$\phi _{\varepsilon}$ by integrating Eq.~\eqref{eq:dBdcosdphi} over
$\cos \theta _{\varepsilon}$, giving
%
\begin{align}
\label{eq:dBdphi}
\frac{dB_{\mu \to 3e}}{d\phi _{\varepsilon}} & =
\frac{B_{\mu \to 3e}}{2\pi}\bigg[1+\frac{\pi}{4}P\big(A^{\text{P}'}_{
\mu \to 3e}\cos \phi _{\varepsilon}+A^{\text{T}}_{\mu \to 3e}\sin
\phi _{\varepsilon}\big)\bigg]
\\
&\equiv \frac{B_{\mu \to 3e}}{2\pi}\Big[1+A^{\text{P}'\text{T}}_{\mu
\to 3e}P\cos (\phi _{\varepsilon}+\omega ^{\text{P}'\text{T}}_{\mu \to 3e})
\Big]\,,
\end{align}
with
$A^{\text{P}'\text{T}}_{\mu \to 3e} = \frac{\pi}{4}\,\text{sgn}(A^{
\text{P}'}_{\mu \to 3e})\sqrt{(A^{\text{P}'}_{\mu \to 3e})^{2}+(A^{
\text{T}}_{\mu \to 3e})^{2}}$ and
$\omega ^{\text{P}'\text{T}}_{\mu \to 3e} = \tan ^{-1}(-A^{\text{T}}_{
\mu \to 3e}/A^{\text{P}'}_{\mu \to 3e})$. Here, the P-odd asymmetry
$A^{\text{P}'}_{\mu \to 3e}$ is given by
%
\begin{align}
\label{eq:APprime}
A^{\text{P}'}_{\mu \to 3e} = \frac{1}{B_{\mu \to 3e}}\frac{128}{35}
\bigg[&\frac{1}{12}|C^{ee}_{V,LR}|^{2} - 21|e C_{D,L}|^{2}
\nonumber
\\
&\, +\,\text{Re}\big[eC_{D,L}(3C^{ee*}_{V,LL}-2C^{ee*}_{V,LR})\big] - (L
\leftrightarrow R)\bigg]\,,
\end{align}
and the T-odd asymmetry $A^{\text{T}}_{\mu \to 3e}$ by
%
\begin{align}
\label{eq:AT}
A^{\text{T}}_{\mu \to 3e} = \frac{1}{B_{\mu \to 3e}}\frac{128}{35}
\bigg[&-\,\text{Im}\big[eC_{D,L}(3C^{ee*}_{V,LL}-2C^{ee*}_{V,LR})\big] +
(L\leftrightarrow R)\bigg]\,.
\end{align}
Fewer operators in Eq.~\eqref{eq:operators} contribute to these asymmetries
at leading order in $m_{e}/m_{\mu}$. In Table~\ref{tab:AAp}, we show the
values of the asymmetry $A^{\text{P}'}_{\mu \to 3e}$ when each of the coefficients
in Eq.~\eqref{eq:operators} is taken to be non-zero at a time. In isolation,
the scalar and vector operators with coefficients $C^{ee}_{S,LL}$,
$C^{ee}_{S,RR}$, $C^{ee}_{V,LL}$ and $C^{ee}_{V,RR}$ do not contribute
to either $A^{\text{P}'}_{\mu \to 3e}$ or $A^{\text{T}}_{\mu \to 3e}$, while
only the imaginary part of the interference between dipole and vector operators
contributes to $A^{\text{T}}_{\mu \to 3e}$.

The general expressions for the three asymmetries are compatible with those
derived in Ref.~\cite{Okada:1999zk}. In contrast, our asymmetry
$A^{\text{P}}_{\mu \to 3e}$ does not depend on the cut-off parameter
$\delta $ which was set to $\delta = 0.02$ in the numerical estimates in~\cite{Okada:1999zk}.
In our case, the role of this parameter is played by
$4m_{e}^{2}/m_{\mu}^{2}$. In addition, the expressions for our asymmetries
are a factor of two larger than those reported in Ref.~\cite{Okada:1999zk}
as a consequence of the way we define them. Further, our T-odd asymmetry
is of opposite sign.

In order to observe CP violation in the
$\mu ^{+}\to e^{+}e^{+}e^{-}$ decay, both the dipole and vector operators
must be present; additionally, the coefficients of these operators must
have a non-zero relative phase. This is possible, for example, in the canonical
type~I seesaw~\cite{Ilakovac:1994kj,Alonso:2012ji,Dinh:2012bp} and general
SUSY~\cite{Okada:1997fz,Ilakovac:2012sh} extensions. In the former, the
dipole coefficient $C_{D,L}$ is generated by a photon-penguin diagram with
a heavy neutral lepton and $W^{\pm}$ boson in the loop. The vector coefficient
$C^{ee}_{V,LL}$, on the other hand, gets contributions from box diagrams
containing two heavy neutral leptons and $W^{\pm}$ bosons. The penguin
and box diagram amplitudes are proportional to
$U_{e i}U^{*}_{\mu j}$ and $U_{e i}U^{*}_{\mu j}U_{e i}U^{*}_{e j}$, respectively,
where $U_{\ell i}$ is the mixing between the charged lepton $\ell $ and
neutrino mass eigenstate $N_{i}$ ($i = 1,\cdots , 3+n$, with $n$ the number
of heavy neutral leptons). The dependence of the coefficients
$C_{D,L}$ and $C^{ee}_{V,LL}$ on additional Dirac CP phases in the extended
mixing matrix can therefore be different. In SUSY extensions, the coefficients
can also depend differently on phases in the soft SUSY-breaking sector.

\section{Phenomenological Analysis}
\label{sec:disc}

We now discuss the implications of a detection of the
$\mu ^{+}\to e^{+}e^{+}e^{-}$ decay of a polarised $\mu ^{+}$. We consider,
in particular, how angular and helicity measurements of the outgoing positrons
and electron can help to identify the NP contributing to the cLFV process.

In future, the $\mu ^{+}\to e^{+}e^{+}e^{-}$ decay will be searched for
by the Mu3e experiment at PSI. Phase~I of the experiment will receive
$\mathcal{O}(10^{8})$ positive muons per second with average momenta
$p\sim 29.8$~MeV from pion decays at rest near the surface of the production
target (the $\pi \text{E}5$ compact muon beamline). The muons produced in
this way are fully polarised; the beam is thus estimated to have an average
polarisation of $P\sim 93\%$. This phase of the experiment is expected
to probe down to $B_{\mu \to 3e} \sim 2\times 10^{-15}$, in contrast to
the current limit of $B_{\mu \to 3e} < 1.0\times 10^{-12}$ from the SINDRUM
experiment~\cite{SINDRUM:1987nra}. This is a much larger improvement on
the current experimental bound compared to that expected with the planned
search for the $\mu ^{+}\to e^{+}\gamma $ decay. Using the
$\pi \text{E}5$ beamline, the MEG II experiment aims to reach
$B_{\mu \to e\gamma} \sim 6\times 10^{-14}$ compared the current bound
of $B_{\mu \to e\gamma} < 4.2\times 10^{-13}$ from the original MEG experiment.
In the longer term, the HIMB proposal aims to provide
$\mathcal{O}(10^{9})$ muons per second to Phase~II of the Mu3e experiment,
allowing to reach even higher sensitivity to $B_{\mu \to 3e}$. The Mu3e
detector is designed to optimally track the positrons and electron from
decays of muons at rest, and therefore to fully reconstruct the event kinematics.
To cover a large portion of the final state phase space, the experiment
aims to reconstruct tracks over a large solid angle (limited by the beam
entry and exit points) with momenta ranging from $\sim 1$~MeV to half the
muon mass~\cite{Mu3e:2020gyw}. The experiment must be able to discriminate the $\mu ^{+}\to e^{+}e^{+}e^{-}$ decay from backgrounds such as
$\mu ^{+}\to e^{+}e^{+}e^{-}\nu \bar{\nu}$ and a Michel positron accompanied
by a $e^{+}e^{-}$ pair from photon conversion or Bhabha scattering. The
experiment can nevertheless exploit the kinematics of the
$\mu ^{+}\to e^{+}e^{+}e^{-}$ decay at rest,
$|\sum _{i} \vec{k}_{i}| = 0$ and $\sum _{i} E_{k_{i}} = m_{\mu}$, to mitigate
these backgrounds.

To understand how the asymmetries can be measured by an experiment such
as Mu3e (in the scenario where $\mu ^{+}\to e^{+}e^{+}e^{-}$ is observed),
we note that they can be extracted from Eq.~\eqref{eq:dBdcosdphi} as follows:
%
\begin{align}
\label{eq:AP2}
A^{\text{P}}_{\mu \to 3e} &=\frac{2}{P}\int ^{2\pi}_{0} d\phi _{
\varepsilon}\bigg(\int ^{1}_{0}-\int ^{0}_{-1}\bigg)d\cos \theta _{
\varepsilon}\,\frac{1}{B_{\mu \rightarrow 3e}}
\frac{dB_{\mu \to 3e}}{d\Omega _{\varepsilon}}\,,
\\
\label{eq:APprime2}
A^{\text{P}'}_{\mu \to 3e} & = \frac{2}{P}\bigg(\int ^{\frac{\pi}{2}}_{-
\frac{\pi}{2}}-\int ^{\frac{3\pi}{2}}_{\frac{\pi}{2}}\bigg)d\phi _{
\varepsilon}\int ^{1}_{-1} d\cos \theta _{\varepsilon}\,
\frac{1}{B_{\mu \rightarrow 3e}}
\frac{dB_{\mu \to 3e}}{d\Omega _{\varepsilon}}\,,
\\
\label{eq:AT2}
A^{\text{T}}_{\mu \to 3e} & = \frac{2}{P}\bigg(\int ^{\pi}_{0}-\int ^{0}_{-
\pi}\bigg)d\phi _{\varepsilon}\int ^{1}_{-1} d\cos \theta _{
\varepsilon}\,\frac{1}{B_{\mu \rightarrow 3e}}
\frac{dB_{\mu \to 3e}}{d\Omega _{\varepsilon}}\,.
\end{align}
The P-odd asymmetry $A^{\text{P}}_{\mu \to 3e}$ can be determined by counting
the number of electrons with outgoing angles $\theta _{\varepsilon}$ in
the range $[0,\frac{\pi}{2}]$ minus those in the range
$[\frac{\pi}{2},\pi ]$. Other than the outgoing electron direction, this
only requires knowledge of the magnitude and direction of the muon polarisation
$\vec{P}$. To instead determine the P-odd and T-odd asymmetries
$A^{\text{P}'}_{\mu \to 3e}$ and $A^{\text{T}}_{\mu \to 3e}$, it is also
necessary to measure the directions of the outgoing positrons to identify
the decay plane and the azimuthal angle $\phi _{\varepsilon}$. The asymmetry
$A^{\text{P}'}_{\mu \to 3e}$ can be ascertained by counting the number of
decay topologies with angles $\phi _{\varepsilon}$ in the range
$[-\frac{\pi}{2},\frac{\pi}{2}]$ minus those in the range
$[\frac{\pi}{2},\frac{3\pi}{2}]$. Conversely, for the asymmetry
$A^{\text{T}}_{\mu \to 3e}$ one counts the number of decays with angles
$\phi _{\varepsilon}$ in the range $[0,\pi ]$ minus those in the range
$[-\pi ,0]$.

In Table~\ref{tab:AAp} we have already shown the values of the asymmetries
$A^{\text{P}}_{\mu \to 3e}$ and $A^{\text{P}'}_{\mu \to 3e}$ when each effective
coefficient is taken to be non-zero at a time. It follows from the results
in Table~\ref{tab:AAp} that the measurement of
$A^{\text{P}}_{\mu \to 3e}$ can allow to discriminate between the dipole,
scalar $RR$ (scalar $LL$), vector $LL$ (vector $RR$), vector $LR$ and vector
$RL$ operators, while it will be impossible to distinguish between scalar
$RR$ (scalar $LL$) and vector $RR$ (vector $LL$) operators. With the measurement
of $A^{\text{P}'}_{\mu \to 3e}$ alone it may be possible to discriminate
between the dipole, scalar or vector $LL$ or $RR$, and vector $LR$ or
$RL$ operators. At the same time, the difference between the values of
$A^{\text{P}'}_{\mu \to 3e}$ due to the dipole $R$ (dipole $L$) and vector
$RL$ (vector $LR$) operators is very small and thus distinguishing between
these operators requires an extremely precise measurement
of $A^{\text{P}'}_{\mu \to 3e}$.

If more than one operator is present, the asymmetries instead take a range
of values depending on the relative sizes of the respective coefficients.
For example, in the presence of the vector coefficients
$C^{ee}_{V,LL}$ and $C^{ee}_{V,LR}$ (with all other coefficients set to
zero), we have
%
\begin{align}
A^{\text{P}}_{\mu \to 3e} &= -1+\frac{4}{3(1+2x)}\,,~~A^{\text{P}'}_{
\mu \to 3e}= \frac{32}{105(1+2x)}\,;~~x\equiv
\frac{|C^{ee}_{V,LL}|^{2}}{|C^{ee}_{V,LR}|^{2}}\,,
\end{align}
which clearly range from the minimum values
$A^{\text{P}-}_{\mu \to 3e} = -1$ and
$A^{\text{P}'-}_{\mu \to 3e} = 0$ when $x \gg 1$ and the maximum values
$A^{\text{P}+}_{\mu \to 3e} = \frac{1}{3}$ and
$A^{\text{P}'+}_{\mu \to 3e} = \frac{32}{105}$ when $x\ll 1$. Actually,
the values of the asymmetries $A^{\text{P}}_{\mu \to 3e}$ and
$A^{\text{P}'}_{\mu \to 3e}$ are correlated:
$A^{\text{P}'}_{\mu \to 3e} = \frac{8}{35}(1+A^{\text{P}}_{\mu \to 3e})$.
Observing this correlation would strongly imply
that the combination of the vector $LL$ and $LR$ operators triggers
the $\mu^{+}\rightarrow e^{+}e^{+}e^{-}$ decay.

\begin{table}[t!]
\centering
\renewcommand{\arraystretch}{1.4}
\setlength\tabcolsep{5.2pt}
\begin{tabular}{|c|c|}
\hline
$C^{(\bar{5})}_i$/$C^{(\bar{6})}_i$  &  Final state helicities \\  \hline\hline
$C_{D,L(R)}$ & $\mu^{+}\to e_{R(L)}^{+}e_{R/L}^{+}e_{L/R}^{-}$ \\
$C^{ee}_{S,LL(RR)}$ & $\mu^{+}\to e_{R(L)}^{+}e_{R(L)}^{+}e_{R(L)}^{-}$ \\
$C^{ee}_{V,LL(RR)}$ & $\mu^{+}\to e_{R(L)}^{+}e_{R(L)}^{+}e_{L(R)}^{-}$ \\
$C^{ee}_{V,LR(RL)}$ & $\mu^{+}\to e_{R(L)}^{+}e_{L(R)}^{+}e_{R(L)}^{-}$ \\\hline
\end{tabular}
\caption{Helicities of the outgoing positrons and electron when the $\mu^{+}\to e^{+} e^{+} e^{-}$ decay is induced by each of the operators in Eq.~\eqref{eq:operators}.}
\label{tab:helicities}
\end{table}

A measurement of the maximal asymmetry
$A^{\text{P}}_{\mu \to 3e} = 1~(-1)$ would imply that one of the coefficients
$C^{ee}_{S,LL}$ or $C^{ee}_{V,RR}$ ($C^{ee}_{S,RR}$ or
$C^{ee}_{V,LL}$) dominates over the others. The other P-odd asymmetry
$A^{\text{P}'}_{\mu \to 3e}$ is approximately zero when both
$C^{ee}_{S,LL}$ and $C^{ee}_{V,RR}$ dominate, while the T-odd asymmetry
(which relies on the interference between the dipole and vector operators)
is also suppressed; neither can therefore be used to distinguish the two
contributions. However, the scalar and vector operators can be distinguished
with an additional measurement of the helicity of the outgoing positrons.
In Table~\ref{tab:helicities}, we show the helicities of the final states
for each of the operators in Eq.~\eqref{eq:operators}. For both the scalar
and vector operators (with the coefficients $C^{ee}_{S,LL}$ and
$C^{ee}_{V,RR}$, respectively) the helicity of the outgoing electron is
positive. The outgoing positrons, on the other hand, have positive helicities
for the scalar operator and negative helicities for the vector operator.
A measurement of these helicities can therefore be used to distinguish
the two scenarios.

\begin{figure*}
\centering
\includegraphics[width=0.45\textwidth]{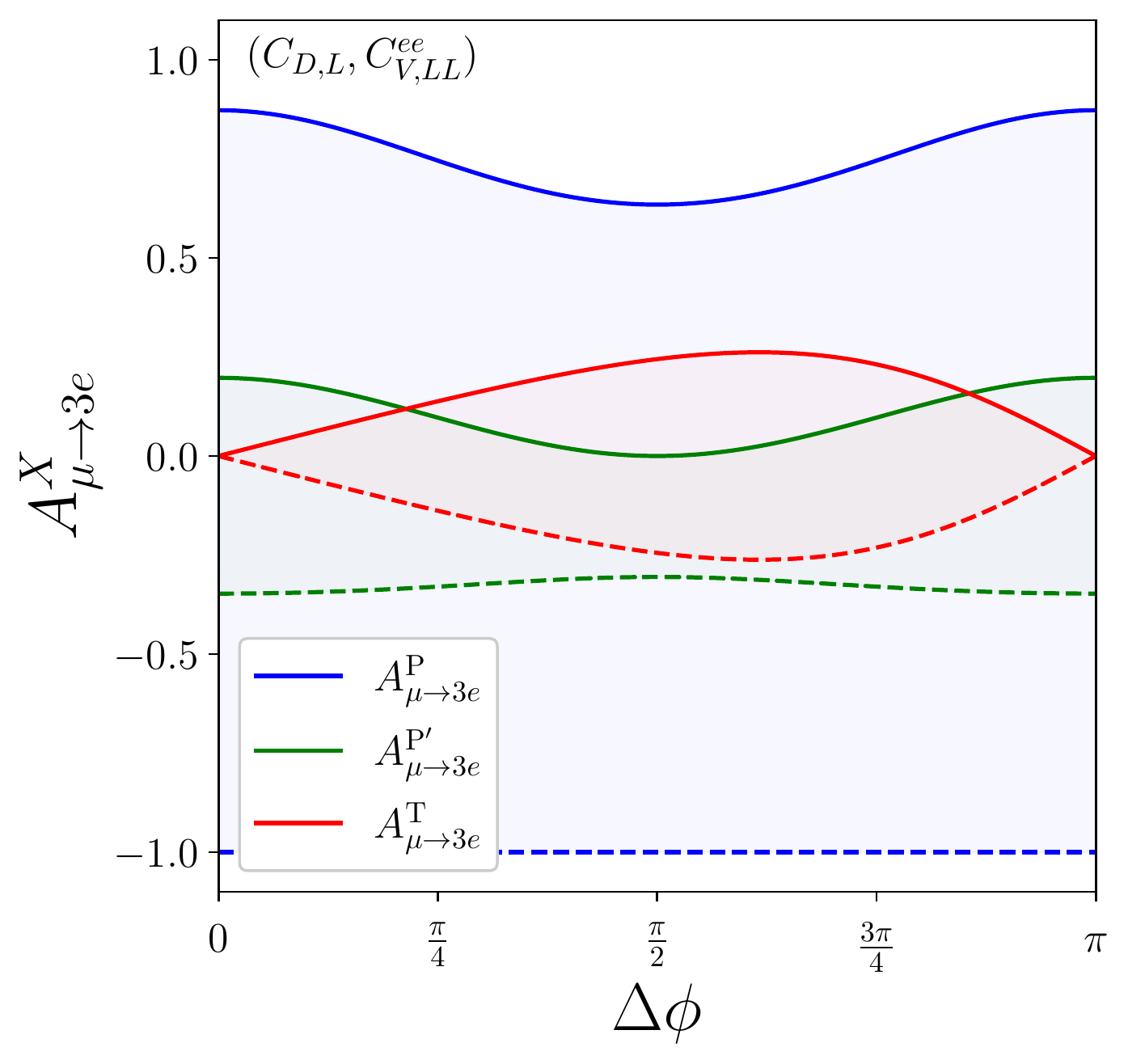}
\includegraphics[width=0.45\textwidth]{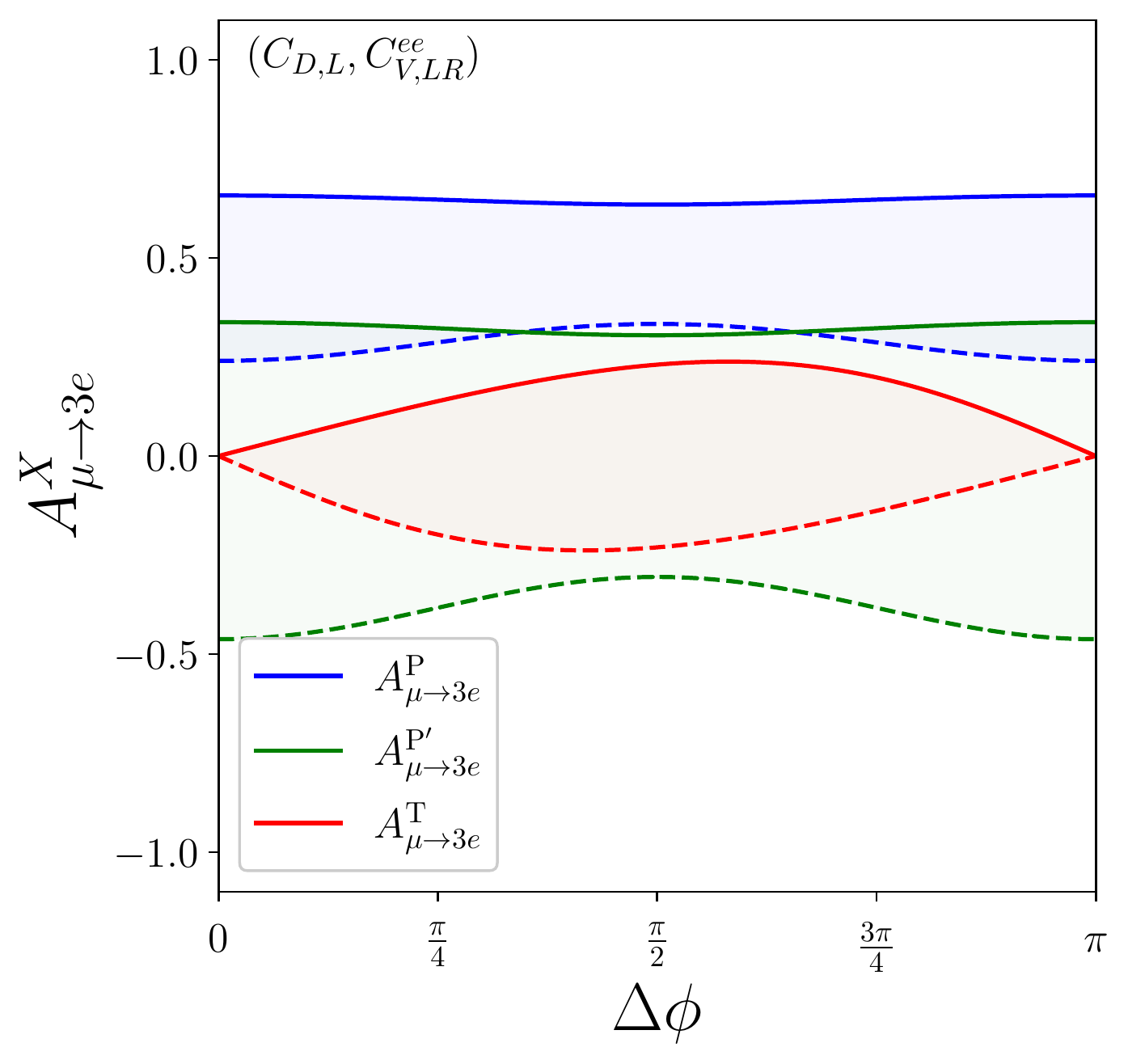}
\caption{Ranges of the asymmetries $A^{\text{P}}_{\mu \to 3e}$ (blue),
$A^{\text{P}'}_{\mu \to 3e}$ (green), and $A^{\text{T}}_{\mu \to 3e}$ (red),
with solid lines indicating the maximum asymmetries $A^{X+}_{\mu \to 3e}$ and
the dashed lines the minimum asymmetries $A^{X-}_{\mu \to 3e}$. In the left
panel, only the coefficients $C_{D,L}$ and $C^{ee}_{V,LL}$ are taken to be
non-zero, while in the right panel only $C_{D,L}$ and $C^{ee}_{V,LR}$ are
considered. The ranges are shown as a function of the phase difference $\Delta
\phi $ between the coefficients.}
\label{fig:minmaxplot}
\end{figure*}

In the presence of the dipole and vector operators, which interfere at
leading order in $\hat{m}_{e}$, the asymmetry does not simply range between
the values in Table~\ref{tab:AAp}. For example, if only the coefficients
$C_{D,L}$ and $C^{ee}_{V,LL}$ are non-zero, the asymmetry does not just
range between $A^{\text{P}-}_{\mu \to 3e} = -1$ and
$A^{\text{P}+}_{\mu \to 3e} = 0.63$ in the limits
$|C_{D,L}|\ll |C^{ee}_{V,LL}|$ and $|C_{D,L}|\gg |C^{ee}_{V,LL}|$, respectively.
Due to the interference between operators, the limits instead depend on
the relative phase $\Delta \phi $ between the coefficients. For the coefficients
$C_{D,L}$ and $C^{ee}_{V,LL}$, the minimum and maximum values of
$A^{\text{P}}_{\mu \to 3e}$ depend on the phase as
%
\begin{align}
\label{eq:APminmax}
A^{\text{P}-}_{\mu \to 3e} &= -1\,,~~ A^{\text{P}+}_{\mu \to 3e} = 1 -
\frac{3\big(57 \ln 2-47 + 4 \cos (2\Delta \phi )\big)}{23 + 6 \cos (2 \Delta \phi ) + 12 \ln (4\hat{m}_{e}^{2})}
\,.
\end{align}
Correspondingly, the minimum and maximum values for
$A^{\text{P}'}_{\mu \to 3e}$ and $A^{\text{T}}_{\mu \to 3e}$ are
%
\begin{align}
\label{eq:APpminmax}
A^{\text{P}'\pm}_{\mu \to 3e} &=
\frac{24\big(15 - 6 \cos (2 \Delta \phi ) \mp \sqrt{
441 - 12(59 +
12 \ln (4\hat{m}_{e}^{2}))\cos ^{2}\Delta \phi}\big)}{35\big(23 +
6 \cos (2 \Delta \phi ) + 12 \ln (4\hat{m}_{e}^{2})\big)} \,,
\\
A^{\text{T}\pm}_{\mu \to 3e} &=
\frac{\pm 144 \sin \Delta \phi}{35 \big(6 \cos \Delta \phi +
\sqrt{-51 - 36 \ln (4\hat{m}_{e}^{2})}\big)}\,.
\label{eq:ATminmax}
\end{align}
In the left panel of Fig.~\ref{fig:minmaxplot}, we depict the ranges of
the asymmetries $A^{\text{P}}_{\mu \to 3e}$ (blue),
$A^{\text{P}'}_{\mu \to 3e}$ (green), and $A^{\text{T}}_{\mu \to 3e}$ (red)
as a function of the relative phase $\Delta \phi $ when only
$C_{D,L}$ and $C^{ee}_{V,LL}$ are taken to be non-zero. The solid and dashed
lines indicate the maximum and minimum asymmetries
$A^{X+}_{\mu \to 3e}$ and $A^{X-}_{\mu \to 3e}$, respectively. It can be
seen that for $\Delta \phi = 0$, the P-odd asymmetry has the wider range
of $-1\leq A^{\text{P}}_{\mu \to 3e} \leq 0.87$ compared to
$-1\leq A^{\text{P}}_{\mu \to 3e} \leq 0.63$ for
$\Delta \phi = \frac{\pi}{2}$. In Fig.~\ref{fig:minmaxplot}, right panel,
we also show the ranges when $C_{D,L}$ and $C^{ee}_{V,LR}$ are taken to
be non-zero.

\begin{figure*}
\centering
\includegraphics[width=0.85\textwidth]{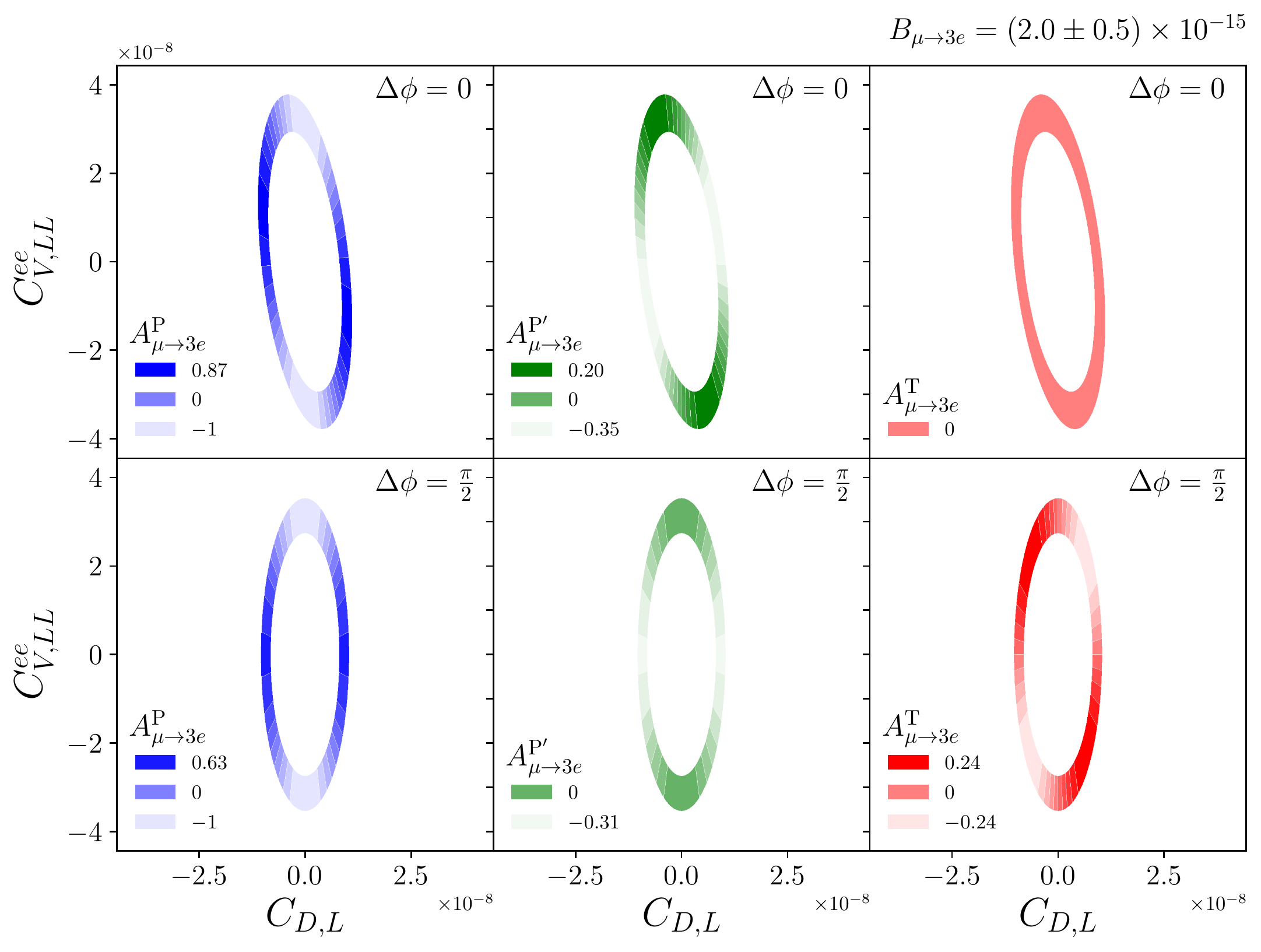}
\caption{Assuming that the cLFV decay $\mu ^{+}\to e^{+}e^{+}e^{-}$ is observed
with a branching ratio of $B_{\mu \to 3e} = (2.0\pm 0.5)\times
10^{-15}$, we depict the allowed regions in the $(C_{D,L}, C^{ee}_{V,LL})$
parameter space for a relative phase difference between the coefficients of
$\Delta \phi =0$ (top) and $\Delta \phi =\frac{\pi}{2}$ (bottom). We also show
the size of the asymmetries $A^{\text{P}}_{\mu \to 3e}$ (left),
$A^{\text{P}'}_{\mu \to 3e}$ (middle), and $A^{\text{T}}_{\mu \to 3e}$ (right).
Measurements of these asymmetries can further constrain the parameter space, or
indicate additional non-zero effective coefficients.}
\label{fig:eta_plot}
\end{figure*}

If a future experiment does not observe
$\mu ^{+}\to e^{+}e^{+}e^{-}$ decay, this sets an upper limit on the branching
ratio $B_{\mu \to e\gamma}$ in Eq.~\eqref{eq:Br} and therefore on each
of the coefficients in Eq.~\eqref{eq:operators} if the other coefficients
are set to zero. A positive detection of
$\mu ^{+}\to e^{+}e^{+}e^{-}$ decay instead sets an allowed range for each
coefficient. For example, an observed branching ratio of
$B_{\mu \to 3e} = (2.0\pm 0.5)\times 10^{-15}$ constrains the dipole coefficients
to lie in the range
$8.1\times 10^{-9}\leq |C_{D,L(R)}|\leq 1.0\times 10^{-8}$. If two coefficients
are allowed to be non-zero, the favoured region becomes an circle or ellipse
in the parameter space of the coefficients. If the dipole and vector coefficients
$C_{D,L}$ and $C^{ee}_{V,LL}$ (or $C^{ee}_{V,LR}$) are taken to be non-zero,
the alignment of the ellipse is controlled by the phase difference
$\Delta \phi $ between the coefficients. In Fig.~\ref{fig:eta_plot}, we
plot the allowed regions in the $(C_{D,L},C^{ee}_{V,LL})$ plane if a future
experiment measures $B_{\mu \to 3e} = (2.0\pm 0.5)\times 10^{-15}$, for
$\Delta \phi = 0$ (top panels) and $\Delta \phi = \frac{\pi}{2}$ (bottom
panels). For $\Delta \phi = 0$, the real part of
$C_{D,L}C^{ee*}_{V,LL}$ in Eq.~\eqref{eq:Br} is non-zero and therefore
the ellipse is not aligned along the $C_{D,L}$ and $C^{ee}_{V,LL}$ axes.
For $\Delta \phi = \frac{\pi}{2}$, the real part is indeed zero and the
ellipse is aligned. In Fig.~\ref{fig:eta_plot}, we allow for negative values
of the coefficients $C_{D,L}$ and $C^{ee}_{V,LL}$. The allowed region for
the phase difference $\Delta \phi = \pi $ is therefore the mirror of the
$\Delta \phi = 0$ ellipse along either the $C_{D,L}$ or
$C^{ee}_{V,LL}$ axis. In the left, centre and right panels of Fig.~\ref{fig:eta_plot}
we also show the magnitudes of the asymmetries
$A^{\text{P}}_{\mu \to 3e}$ (blue), $A^{\text{P}'}_{\mu \to 3e}$ (green),
and $A^{\text{T}}_{\mu \to 3e}$ (red), respectively. The ranges of these
asymmetries are consistent with the minimum and maximum values in Eqs.~\eqref{eq:APminmax},
\eqref{eq:APpminmax} and \eqref{eq:ATminmax}.

We would finally like to note that a measurement of the process
$\mu ^{-}e^{-}\to e^{-}e^{-}$ in a muonic atom, first suggested in Ref.~\cite{Koike:2010xr},
can provide additional information on the effective coefficients of Eq.~\eqref{eq:operators}.
Assuming a non-relativistic bound state for the muon, a 1S electron, a
point nuclear charge and the plane wave approximation for the final state
electrons, one obtains an analytical result for the rate
$\Gamma (\mu ^{-}e^{-}\to e^{-}e^{-})$ with interference terms between
the four-fermion operators~\cite{Uesaka:2016vfy}. The dependence of the
rate on the dipole coefficients was studied separately in Ref.~\cite{Uesaka:2017yin};
however, in principle there are also interference terms between the dipole
and four-fermion operators. In a similar manner to the
$\mu ^{+}\to e^{+}e^{+}e^{-}$ decay, the angular distribution of the
outgoing electrons for a polarised muon can further identify the contribution
of each operator~\cite{Kuno:2019ttl}. In future, it is possible for Phase~I
of the COMET experiment to search for the clean signal of two back-to-back
electrons with a total energy of the muon mass minus the muon binding energy~\cite{COMET:2018auw}.
A complete study of the complementarity of the cLFV processes in Table~\ref{tab:mutoeprocesses}
(including measurements of angular distributions and helicities) is beyond
the scope of this work.

\section{Conclusions}
\label{sec:conc}

In the present article, following a purely phenomenological approach, we
have investigated the possibility to discriminate between different effective
operators contributing at tree level to the amplitude of the polarised
$\mu ^{+}\to e^{+}e^{+}e^{-}$ decay by using data on the three angular
observables -- two P-odd and one T-odd asymmetries,
$A^{\text{P}}_{\mu \to 3e}$, $A^{\text{P}'}_{\mu \to 3e}$ and
$A^{\text{T}}_{\mu \to 3e}$, defined in Eqs.~\eqref{eq:AP2},
\eqref{eq:APprime2} and \eqref{eq:AT2}. The consideration of these angular
observables is particularly relevant as the next generation of experiments
on $\mu \leftrightarrow e$ cLFV processes are likely to use polarised muon
beams. In particular, the high-intensity muon beams (HIMB) proposal at
PSI aims to provide a flux of $\mathcal{O}(10^{9})$ longitudinally-polarised
muons per second to the Mu3e and MEG II experiments~\cite{Aiba:2021bxe}
on $\mu ^{+}\to e^{+}e^{+}e^{-}$ and $\mu ^{+}\to e^{+}\gamma $ decays,
with the average $\mu ^{+}$ polarisation estimated to be approximately
of 93\%. The measurement of the P-odd asymmetry
$A^{\text{P}}_{\mu \to 3e}$ in the decay
$\mu ^{+}\to e^{+}e^{+}e^{-}$ requires the knowledge only of the direction
of the outgoing electron and of the magnitude and the direction of the
$\mu ^{+}$ polarisation. To determine the P-odd and T-odd asymmetries
$A^{\text{P}'}_{\mu \to 3e}$ and $A^{\text{T}}_{\mu \to 3e}$, it is also
necessary to measure the directions of the outgoing positrons to identify
the decay plane and the azimuthal angle $\phi _{\varepsilon}$ between the
$\mu ^{+}$ polarisation vector and the decay plane. The effective operators
considered by us respect the CPT symmetry and thus an observation of a
non-zero T-odd asymmetry would imply violation of the CP symmetry. This
is only possible if at least two operators, which have different phases
and whose interference term in the decay rate is not zero (i.e., is not
strongly suppressed), generate the $\mu ^{+}\to e^{+}e^{+}e^{-}$ decay.

The aim of our study was not to perform a comprehensive analysis of all
operators that can induce the $\mu ^{+}\to e^{+}e^{+}e^{-}$ decay, but
to illustrate the effectiveness of the method on the example of the subset
of eight commonly-used effective operators that generate the decay at tree
level, given in Eq. \eqref{eq:operators}. This subset includes dimension-five
left-handed ($L$) and right-handed ($R$) dipole operators and dimension-six
four-fermion operators involving products of scalar and vector currents
of different chiralities (left-left ($LL$), right-right ($RR$), left-right
($LR$) and right-left ($RL$)). The only operators, among those considered
by us, that were found to have significant interference terms in the
$\mu ^{+}\to e^{+}e^{+}e^{-}$ decay rate at lowest order in
$m^{2}_{e}/m^{2}_{\mu}$, are the dipole and vector operators. The interference
between the other operators was therefore neglected. Assuming the scale of
NP to be around $\Lambda _{\text{NP}}\sim 1$~TeV, one can neglect to a good
approximation in the analysed P-odd and T-odd asymmetries the effects of
the renormalisation group running of the considered operators above and
below the electroweak scale.

It follows from the results of our study that if only one of the set of
considered operators triggers the $\mu ^{+}\to e^{+}e^{+}e^{-}$ decay,
the measurement of the P-odd asymmetry $A^{\text{P}}_{\mu \to 3e}$ can allow
to discriminate between the dipole, scalar $RR$ (scalar $LL$), vector
$LL$ (vector $RR$), vector $LR$ and vector $RL$ operators, while it will
be impossible to distinguish the scalar $RR$ (scalar $LL$) and vector
$RR$ (vector $LL$) operators (cf.~Table~\ref{tab:AAp}). With the measurement
of $A^{\text{P}'}_{\mu \to 3e}$ alone it may be possible to discriminate
between the dipole, scalar or vector $LL$ or $RR$, and vector $LR$ or
$RL$ operators. At the same time, the difference between the values of
$A^{\text{P}'}_{\mu \to 3e}$ due to the dipole $R$ (dipole $L$) and vector
$RL$ (vector $LR$) operators is very small (Table~\ref{tab:AAp}) and thus
distinguishing between these operators requires an extremely high precision
in the measurement of $A^{\text{P}'}_{\mu \to 3e}$.
In addition to the case of having just one operator responsible for the
decay, we have derived predictions for the asymmetries of interest also
in the cases when two operators are triggering the decay (Section~\ref{sec:disc}).
Some of these predictions are illustrated in Figs.~\ref{fig:minmaxplot}
and \ref{fig:eta_plot}. The observation of a non-zero T-odd asymmetry in
the polarised $\mu ^{+}\to e^{+}e^{+}e^{-}$ decay, for example, would imply
that, within the considered set, dipole and vector operators with different phases are generating the decay.

Our study has demonstrated that if the $\mu \leftrightarrow e$ cLFV processes are observed, measuring the angular distributions and the corresponding P-odd and T-odd asymmetries
of the $\mu ^{+}\to e^{+}e^{+}e^{-}$ decay with a polarised $\mu ^{+}$ would
provide an additional effective method of identifying the beyond the SM
physics operators that trigger these processes.

\section*{Acknowledgments}

P.~D.~B. has received support from the European Union's Horizon 2020 research and innovation programme under the Marie Sk\l{}odowska-Curie grant agreement No 860881-HIDDeN. The work of S. T. P. was supported in part by the European Union's Horizon 2020 research and innovation programme under the Marie Sk\l{}odowska-Curie grant agreement No.~860881-HIDDeN, by the Italian INFN program on Theoretical Astroparticle Physics 
and by the World Premier International Research Center Initiative (WPI Initiative, MEXT), Japan.

\newpage

\appendix

{\noindent\bf\Large Appendix}

\section{Relevant Functions}
\label{app:functions}

The functions $C_{i}(x_1,x_2)$ appearing in the differential branching ratio for the $\mu^{+}\to e^{+}e^{+}e^{-}$ decay in Eq.~\eqref{eq:diffB}, where $x_1 \equiv 2E_{k_1}/m_\mu$ and $x_2 \equiv 2E_{k_2}/m_\mu$ and $E_{k_1}\geq E_{k_1}$ are the outgoing positron energies, are of the form
\begin{align}
C_1(x_1,x_2) =&\,\, \bigg(\frac{|C^{ee}_{S,LL}|^2}{16}+|C^{ee}_{V,LL}|^2\bigg)F_{1}(x_1,x_2) \nonumber\\
&+|C^{ee}_{V,LR}|^2 F_{2}(x_1,x_2) + |eC_{D,L}|^2 F_{3}(x_1,x_2)\nonumber\\
&+\text{Re}\big[eC_{D,L}C^*_{V,LL}\big]F_{4}(x_1,x_2)+\text{Re}\big[eC_{D,L}C^*_{V,LR}\big]F_{5}(x_1,x_2) \nonumber\\
&+ (L\leftrightarrow R)\,,\\
C_2(x_1,x_2) =&\,\, \bigg(\frac{|C^{ee}_{S,LL}|^2}{16}-|C^{ee}_{V,LL}|^2\bigg)F_1(x_1,x_2) \nonumber \\
&-|C^{ee}_{V,LR}|^2 G_2(x_1,x_2)+|eC_{D,L}|^2G_3(x_1,x_2) \nonumber\\
&-\text{Re}\big[eC_{D,L}C^*_{V,LL}\big]F_4(x_1,x_2)+\text{Re}\big[eC_{D,L}C^*_{V,LR}\big]F_5(x_1,x_2) \nonumber\\
& - (L\leftrightarrow R)\,,\\
C_3(x_1,x_2) =
&-|C^{ee}_{V,LR}|^2H_2(x_1,x_2)+|eC_{D,L}|^2H_3(x_1,x_2) \nonumber\\
&+\text{Re}\big[eC_{D,L}C^*_{V,LL}\big]H_4(x_1,x_2) - \text{Re}\big[eC_{D,L}C^*_{V,LR}\big]H_5(x_1,x_2)\nonumber\\
& - (L\leftrightarrow R) \,, \\
C_4(x_1,x_2) =&\,\,
\text{Im}\big[eC_{D,L}C^*_{V,LL}]H_4(x_1,x_2)-\text{Im}\big[eC_{D,L}C^*_{V,LR}\big]H_5(x_1,x_2) \nonumber\\
& + (L\leftrightarrow R) \,.
\end{align}
Here, the functions $F_{i}(x_1,x_2)$ are given by
\begin{align}
F_1(x_1,x_2) &= 8(2-x_1-x_2)(x_1+x_2-1) \,, \nonumber\\
F_2(x_1,x_2) &= 2\left[x_1(1-x_1)+x_2(1-x_2)\right] \,, \nonumber\\
F_3(x_1,x_2) &= 8\left[\frac{2x_1^2-2x_1+1}{1-x_2}+\frac{2x_2^2-2x_2+1}{1-x_1}\right] \,,\nonumber\\
F_4(x_1,x_2) &= 32(x_1+x_2-1) \,,\nonumber\\
F_5(x_1,x_2) &= 8(2-x_1-x_2)\,,
\end{align}
the functions $G_{i}(x_1,x_2)$ by
\begin{align}
G_2(x_1,x_2) &= 2 \frac{(x_1+x_2)(x_1^2+x_2^2)-3(x_1+x_2)^{2}+6(x_1+x_2)-4}{2-x_1-x_2} \,,\nonumber\\
G_3(x_1,x_2) &= \frac{8}{(1-x_1)(1-x_2)(2-x_1-x_2)} \nonumber\\
&\quad\times\left[2(x_1+x_2)(x_1^{3}+x_2^{3})-4(x_1 + x_2)(2 x_1^{2}+x_1 x_2 + 2 x_2^{2})\right. \nonumber\\
&\quad\quad~~\left.+(19 x_1^{2}+30 x_1 x_2 + 19 x_2^{2})-12(2 x_1 + 2 x_2 - 1)\right]\,,
\end{align}
and finally the $H_{i}(x_1,x_2)$ functions by
\begin{align}
H_2(x_1,x_2) &= \frac{4(x_2-x_1)}{2-x_1-x_2}\sqrt{(1-x_1)(1-x_2)(x_1+x_2-1)} \,,\nonumber\\
H_3(x_1,x_2) &= \frac{32(x_2-x_1)(x_1+x_2-1)}{2-x_1-x_2}\sqrt{\frac{x_1+x_2-1}{(1-x_1)(1-x_2)}} \,,\nonumber\\
H_4(x_1,x_2) &= 16(x_2-x_1)(x_1+x_2-1)\sqrt{\frac{x_1+x_2-1}{(1-x_1)(1-x_2)}} \,,\nonumber\\
H_5(x_1,x_2) &= 8(x_2-x_1)(2-x_1-x_2)\sqrt{\frac{x_1+x_2-1}{(1-x_1)(1-x_2)}}\,.
\end{align}
In these expressions we have neglected the electron mass $m_e$. These expressions are in agreement with those in Ref.~\cite{Okada:1999zk}.

It is straightforward to find the single differential branching ratio in the parameter $x_1 = 2E_{k_1}/m_{\mu}$ as
\begin{align}
\frac{dB_{\mu\to 3e}}{dx_1} =&\,\, \bigg(\frac{|C^{ee}_{S,LL}|^2}{16}+|C^{ee}_{V,LL}|^2\bigg)F'_{1}(x_1) \nonumber\\
&+|C^{ee}_{V,LR}|^2 F'_{2}(x_1) + |eC_{D,L}|^2 F'_{3}(x_1)\nonumber\\
&+\text{Re}\big[eC_{D,L}C^*_{V,LL}\big]F'_{4}(x_1)+\text{Re}\big[eC_{D,L}C^*_{V,LR}\big]F'_{5}(x_1) \nonumber\\
&+ (L\leftrightarrow R)\,,
\end{align}
where the $F'_{i}(x_1)$ functions are computed as
\begin{align}
F'_{i}(x_1) = 6\int^{x_1}_{1-x_1} dx_2 \,F_{i}(x_1,x_2)\,.
\end{align}

As mentioned in the main text, for the dipole operators one encounters a logarithmic divergence when neglecting the electron mass and performing the phase-space integral over $x_1$. The approach of Ref.~\cite{Okada:1999zk} was to introduce a cut-off parameter $\delta$ in the upper integration limit of $x_{1}$. However, the physical interpretation of $\delta$ is unclear, and for numerical estimates it must be set arbitrarily to some value. In order to find the exact dependence of the total branching ratio $B_{\mu\rightarrow 3e}$ and the P-odd asymmetry $A^{\text{P}}_{\mu\rightarrow 3e}$ on the electron mass $m_e$, we instead express the differential branching ratio for $\mu^{+}\to e^{+}e^{+}e^{-}$ decay as a function of the kinematic variables $s_1 = (p - k_1)^2 = m_{\mu}^2(1-x_1)+m_e^2$ and $s_2 = (p - k_2)^2 = m_{\mu}^2(1-x_2)+m_e^2$. We find (considering only the dipole operators)
\begin{align}
\frac{dB^{D,L(R)}_{\mu\to 3e}}{ds_1 ds_2 d\Omega_{\varepsilon}} = \frac{3}{2\pi}\Big[&\widehat{C}_1(s_1,s_2) +\widehat{C}_2(s_1,s_2)P \cos\theta_{\varepsilon}+\widehat{C}_3(s_1,s_2)P \sin\theta_{\varepsilon}\cos\phi_{\varepsilon}\Big]\,,
\label{eq:diffB2}
\end{align}
where the functions $\widehat{C}_{i}(s_1,s_2)$ are given by
\begin{align}
\widehat{C}_1(s_1,s_2) =&\,\, \frac{8X}{m_{\mu}^6}\bigg[-3m_{e}^{2}+\frac{m_{\mu}^{4}+3m_{e}^{4}-2 (m_{\mu}^{2}+3m_e^2)s_{1}+2 s_{1}^{2}}{s_2}\nonumber\\
&\quad\quad~\,\,+\frac{2 m_{e}^{2}(m_{\mu}^{2}-m_{e}^{2})^{2}}{s_{2}^{2}}+\frac{m_{e}^{2}(m_{\mu}^2-m_{e}^{2})(m_{\mu}^2-2 m_{e}^{2})}{s_{1}s_{2}} +(s_{1} \leftrightarrow s_{2})\bigg]\,,\\
\widehat{C}_2(s_1,s_2) =&\,\,\frac{8Y}{m_{\mu}^8}\bigg[- 
 3 m_\mu^4 - 2 m_e^2 m_\mu^2 + 9 m_e^4 + 3(m_\mu^2-2 m_e^2)(s_1 + s_2) + s_1^2 + s_2^2\nonumber\\
&\quad\quad~\,\,-\frac{2 m_e^2 m_\mu^2(m_\mu^2-m_e^2) - (m_\mu^4 + 15 m_e^4) s_1 + 2(m_\mu^2 + 5 m_e^2 
    ) s_1^2 - 2 s_1^3}{s_2}\nonumber \\
&\quad\quad~\,\, + \frac{2 m_e^2 (m_\mu^2 - m_e^2)^2 (s_1 - 2 m_e^2)}{s_2^2} \nonumber \\
&\quad\quad~\,\, + \frac{2 m_e^2 (m_\mu^2 - m_e^2)^2 (m_\mu^2 - 2 m_e^2) }{s_1 s_2} + (s_{1} \leftrightarrow s_{2})\bigg] \,,\\
\widehat{C}_3(s_1,s_2) =&\,\, \frac{16Z}{m^4_\mu}\bigg[\frac{2m_\mu^2+m_e^2-2s_1}{s_2}+\frac{2m_e^2(m_\mu^2-m_e^2)}{s_2^2}-(s_{1} \leftrightarrow s_{2})\bigg]\,,
\end{align}
with $X \equiv |eC_{D,L}|^2+|eC_{D,R}|^2$ and
\begin{align}
Y \equiv \frac{1}{\lambda^{\frac{1}{2}}\big(\frac{m_e^2}{m_\mu^2},\frac{s}{m_\mu^2}\big)}\big(|eC_{D,L}|^2-|eC_{D,R}|^2\big)\,,~~Z \equiv \frac{Y}{m_\mu^3}\sqrt{s s_1 s_2-m_e^2(m_\mu^2-m_e^2)^2}\,,
\end{align}
where we have used $\lambda(x,y) = (1-x-y)^2-4xy$ and $s = m_\mu^2+3m_e^2-s_1-s_2$. Using the phase-space integration limits of $s_1$ and $s_2$ in Eqs.~\eqref{eq:s2limits} and \eqref{eq:s3limits}, respectively, we then obtain the total branching ratio as
\begin{align}
B^{D,L(R)}_{\mu\to 3e} &= 6\int^{(m_\mu-m_e)^2}_{4m_e^2}ds_1 \int^{s^{+}_2(s_1)}_{s_1}ds_2 ~\widehat{C}_1(s_1,s_2) \nonumber\\
&= -\frac{8}{3}\big(|eC_{D,L}|^2+|eC_{D,R}|^2\big)L_1(\hat{m}_e^2)\,,
\end{align}
where $L_1(x) = 17+12\ln 4x$ and we have only retained the terms at lowest order in an expansion in $\hat{m}_e = \frac{m_e}{m_{\mu}}$. Similarly, the P-odd
asymmetry is found to be
%
\begin{align}
\big(A^{\text{P}}_{\mu \to 3e}\big)^{D,L(R)} &= \frac{6}{B^{D,L(R)}_{\mu \to 3e}}
\int ^{(m_{\mu}-m_{e})^{2}}_{4m_{e}^{2}}ds_{1} \int ^{s^{+}_{2}(s_{1})}_{s_{1}}ds_{2}
~\widehat{C}_{2}(s_{1},s_{2})
\nonumber
\\
&= \frac{1}{B^{D,L(R)}_{\mu \to 3e}}\bigg[-\frac{8}{3}\big(|eC_{D,L}|^{2}-|eC_{D,R}|^{2}
\big)L_{2}(\hat{m}_{e}^{2})\bigg]\,,
\end{align}
where $L_{2}(x) = 170-171\ln 2 + 12\ln 4x$.

\bibliographystyle{BiblioStyle.bst}
\bibliography{references}

\providecommand{\href}[2]{#2}\begingroup\raggedright\begin{thebibliography}{10}

\bibitem{KATRIN:2019yun}
{\bf KATRIN} Collaboration, M.~Aker {\em et.~al.}, {\it {Improved Upper Limit
  on the Neutrino Mass from a Direct Kinematic Method by KATRIN}},  Phys. Rev.
  Lett. {\bf 123} (2019), no.~22 221802,
  [\href{http://arxiv.org/abs/1909.06048}{{\tt arXiv:1909.06048}}].

\bibitem{Aker:2021gma}
M.~Aker {\em et.~al.}, {\it {First direct neutrino-mass measurement with sub-eV
  sensitivity}},  \href{http://arxiv.org/abs/2105.08533}{{\tt
  arXiv:2105.08533}}.

\bibitem{Planck:2018vyg}
{\bf Planck} Collaboration, N.~Aghanim {\em et.~al.}, {\it {Planck 2018
  results. VI. Cosmological parameters}},  Astron. Astrophys. {\bf 641} (2020)
  A6, [\href{http://arxiv.org/abs/1807.06209}{{\tt arXiv:1807.06209}}].
  [Erratum: Astron.Astrophys. 652, C4 (2021)].

\bibitem{ParticleDataGroup:2020ssz}
{\bf Particle Data Group} Collaboration, P.~A. Zyla {\em et.~al.}, {\it {Review
  of Particle Physics}},  PTEP {\bf 2020} (2020), no.~8 083C01. See Neutrinos
  in Cosmology review by J. Lesgourgues and L. Verde.

\bibitem{Bilenky:1987ty}
S.~M. Bilenky and S.~T. Petcov, {\it {Massive Neutrinos and Neutrino
  Oscillations}},  Rev. Mod. Phys. {\bf 59} (1987) 671. [Erratum: Rev.Mod.Phys.
  61, 169 (1989), Erratum: Rev.Mod.Phys. 60, 575--575 (1988)].

\bibitem{Petcov:1976ff}
S.~T. Petcov, {\it {The Processes $\mu \rightarrow e + \gamma, \mu \rightarrow
  e + \overline{e}, \nu' \rightarrow \nu + \gamma$ in the Weinberg-Salam Model
  with Neutrino Mixing}},  Sov. J. Nucl. Phys. {\bf 25} (1977) 340. [Erratum:
  Sov.J.Nucl.Phys. 25, 698 (1977), Erratum: Yad.Fiz. 25, 1336 (1977)].

\bibitem{Bilenky:1977du}
S.~M. Bilenky, S.~T. Petcov, and B.~Pontecorvo, {\it {Lepton Mixing, mu
  --\ensuremath{>} e + gamma Decay and Neutrino Oscillations}},  Phys. Lett. B
  {\bf 67} (1977) 309.

\bibitem{Cheng:1976uq}
T.~P. Cheng and L.-F. Li, {\it {Nonconservation of Separate mu - Lepton and e -
  Lepton Numbers in Gauge Theories with v+a Currents}},  Phys. Rev. Lett. {\bf
  38} (1977) 381.

\bibitem{MEG:2016leq}
{\bf MEG} Collaboration, A.~M. Baldini {\em et.~al.}, {\it {Search for the
  lepton flavour violating decay $\mu ^+ \rightarrow \mathrm {e}^+ \gamma $
  with the full dataset of the MEG experiment}},  Eur. Phys. J. C {\bf 76}
  (2016), no.~8 434, [\href{http://arxiv.org/abs/1605.05081}{{\tt
  arXiv:1605.05081}}].

\bibitem{Marciano:1977wx}
W.~J. Marciano and A.~I. Sanda, {\it {Exotic Decays of the Muon and Heavy
  Leptons in Gauge Theories}},  Phys. Lett. B {\bf 67} (1977) 303--305.

\bibitem{Petcov:1977ab}
S.~T. Petcov, {\it {Heavy Neutral Lepton Mixing and mu --\ensuremath{>} 3 e
  Decay}},  Phys. Lett. B {\bf 68} (1977) 365--368.

\bibitem{Ilakovac:1994kj}
A.~Ilakovac and A.~Pilaftsis, {\it {Flavor violating charged lepton decays in
  seesaw-type models}},  Nucl. Phys. B {\bf 437} (1995) 491,
  [\href{http://arxiv.org/abs/hep-ph/9403398}{{\tt hep-ph/9403398}}].

\bibitem{Dinh:2012bp}
D.~N. Dinh, A.~Ibarra, E.~Molinaro, and S.~T. Petcov, {\it {The $\mu - e$
  Conversion in Nuclei, $\mu \to e \gamma, \mu \to 3e$ Decays and TeV Scale
  See-Saw Scenarios of Neutrino Mass Generation}},  JHEP {\bf 08} (2012) 125,
  [\href{http://arxiv.org/abs/1205.4671}{{\tt arXiv:1205.4671}}]. [Erratum:
  JHEP 09, 023 (2013)].

\bibitem{Alonso:2012ji}
R.~Alonso, M.~Dhen, M.~B. Gavela, and T.~Hambye, {\it {Muon conversion to
  electron in nuclei in type-I seesaw models}},  JHEP {\bf 01} (2013) 118,
  [\href{http://arxiv.org/abs/1209.2679}{{\tt arXiv:1209.2679}}].

\bibitem{Dinh:2013vya}
D.~N. Dinh and S.~T. Petcov, {\it {Lepton Flavor Violating $\tau$ Decays in TeV
  Scale Type I See-Saw and Higgs Triplet Models}},  JHEP {\bf 09} (2013) 086,
  [\href{http://arxiv.org/abs/1308.4311}{{\tt arXiv:1308.4311}}].

\bibitem{Abada:2018nio}
A.~Abada and A.~M. Teixeira, {\it {Heavy neutral leptons and high-intensity
  observables}},  Front. in Phys. {\bf 6} (2018) 142,
  [\href{http://arxiv.org/abs/1812.08062}{{\tt arXiv:1812.08062}}].

\bibitem{Minkowski:1977sc}
P.~Minkowski, {\it {$\mu \to e\gamma$ at a Rate of One Out of $10^{9}$ Muon
  Decays?}},  Phys. Lett. B {\bf 67} (1977) 421--428.

\bibitem{Yanagida:1979as}
T.~Yanagida, {\it {Horizontal Symmetry And Masses Of Neutrinos}},  Conf.Proc.
  {\bf C7902131} (1979) 95.

\bibitem{GellMann:1980vs}
M.~Gell-Mann, P.~Ramond, and R.~Slansky, {\it {Complex Spinors and Unified
  Theories}},  Conf. Proc. {\bf C790927} (1979) 315--321,
  [\href{http://arxiv.org/abs/1306.4669}{{\tt arXiv:1306.4669}}].

\bibitem{Glashow:1979}
S.~Glashow, {\it Quarks and leptons}, . ed. M. Levy et al. (Plenum, New York),
  p. 707.

\bibitem{mohapatra:1981yp}
R.~N. Mohapatra and G.~Senjanovi\'{c}, {\it Neutrino masses and mixings in
  gauge models with spontaneous parity violation},  Phys. Rev. {\bf D23} (1981)
  165.

\bibitem{Konetschny:1977bn}
W.~Konetschny and W.~Kummer, {\it {Nonconservation of Total Lepton Number with
  Scalar Bosons}},  Phys. Lett. B {\bf 70} (1977) 433--435.

\bibitem{Magg:1980ut}
M.~Magg and C.~Wetterich, {\it {Neutrino Mass Problem and Gauge Hierarchy}},
  Phys. Lett. B {\bf 94} (1980) 61--64.

\bibitem{Cheng:1980qt}
T.~P. Cheng and L.-F. Li, {\it {Neutrino Masses, Mixings and Oscillations in
  SU(2) x U(1) Models of Electroweak Interactions}},  Phys. Rev. D {\bf 22}
  (1980) 2860.

\bibitem{Foot:1988aq}
R.~Foot, H.~Lew, X.~G. He, and G.~C. Joshi, {\it {Seesaw Neutrino Masses
  Induced by a Triplet of Leptons}},  Z. Phys. C {\bf 44} (1989) 441.

\bibitem{Abada:2008ea}
A.~Abada, C.~Biggio, F.~Bonnet, M.~B. Gavela, and T.~Hambye, {\it {mu
  ---\ensuremath{>} e gamma and tau ---\ensuremath{>} l gamma decays in the
  fermion triplet seesaw model}},  Phys. Rev. D {\bf 78} (2008) 033007,
  [\href{http://arxiv.org/abs/0803.0481}{{\tt arXiv:0803.0481}}].

\bibitem{Zee:1980ai}
A.~Zee, {\it {A Theory of Lepton Number Violation, Neutrino Majorana Mass, and
  Oscillation}},  Phys. Lett. B {\bf 93} (1980) 389. [Erratum: Phys.Lett.B 95,
  461 (1980)].

\bibitem{Petcov:1982en}
S.~T. Petcov, {\it {Remarks on the Zee Model of Neutrino Mixing (mu
  ---\ensuremath{>} e gamma, Heavy Neutrino ---\ensuremath{>} Light Neutrino
  gamma, etc.)}},  Phys. Lett. B {\bf 115} (1982) 401--406.

\bibitem{Babu:1988ki}
K.~S. Babu, {\it {Model of 'Calculable' Majorana Neutrino Masses}},  Phys.
  Lett. B {\bf 203} (1988) 132--136.

\bibitem{Ma:2006km}
E.~Ma, {\it {Verifiable radiative seesaw mechanism of neutrino mass and dark
  matter}},  Phys. Rev. D {\bf 73} (2006) 077301,
  [\href{http://arxiv.org/abs/hep-ph/0601225}{{\tt hep-ph/0601225}}].

\bibitem{Cai:2017jrq}
Y.~Cai, J.~Herrero-Garc\'\i{}a, M.~A. Schmidt, A.~Vicente, and R.~R. Volkas,
  {\it {From the trees to the forest: a review of radiative neutrino mass
  models}},  Front. in Phys. {\bf 5} (2017) 63,
  [\href{http://arxiv.org/abs/1706.08524}{{\tt arXiv:1706.08524}}].

\bibitem{Cirigliano:2005ck}
V.~Cirigliano, B.~Grinstein, G.~Isidori, and M.~B. Wise, {\it {Minimal flavor
  violation in the lepton sector}},  Nucl. Phys. B {\bf 728} (2005) 121--134,
  [\href{http://arxiv.org/abs/hep-ph/0507001}{{\tt hep-ph/0507001}}].

\bibitem{Davidson:2006bd}
S.~Davidson and F.~Palorini, {\it {Various definitions of Minimal Flavour
  Violation for Leptons}},  Phys. Lett. B {\bf 642} (2006) 72--80,
  [\href{http://arxiv.org/abs/hep-ph/0607329}{{\tt hep-ph/0607329}}].

\bibitem{Gavela:2009cd}
M.~B. Gavela, T.~Hambye, D.~Hernandez, and P.~Hernandez, {\it {Minimal Flavour
  Seesaw Models}},  JHEP {\bf 09} (2009) 038,
  [\href{http://arxiv.org/abs/0906.1461}{{\tt arXiv:0906.1461}}].

\bibitem{Alonso:2011jd}
R.~Alonso, G.~Isidori, L.~Merlo, L.~A. Munoz, and E.~Nardi, {\it {Minimal
  flavour violation extensions of the seesaw}},  JHEP {\bf 06} (2011) 037,
  [\href{http://arxiv.org/abs/1103.5461}{{\tt arXiv:1103.5461}}].

\bibitem{Dinh:2017smk}
D.~N. Dinh, L.~Merlo, S.~T. Petcov, and R.~Vega-\'Alvarez, {\it {Revisiting
  Minimal Lepton Flavour Violation in the Light of Leptonic CP Violation}},
  JHEP {\bf 07} (2017) 089, [\href{http://arxiv.org/abs/1705.09284}{{\tt
  arXiv:1705.09284}}].

\bibitem{Calibbi:2017uvl}
L.~Calibbi and G.~Signorelli, {\it {Charged Lepton Flavour Violation: An
  Experimental and Theoretical Introduction}},  Riv. Nuovo Cim. {\bf 41}
  (2018), no.~2 71--174, [\href{http://arxiv.org/abs/1709.00294}{{\tt
  arXiv:1709.00294}}].

\bibitem{Gabbiani:1988rb}
F.~Gabbiani and A.~Masiero, {\it {FCNC in Generalized Supersymmetric
  Theories}},  Nucl. Phys. B {\bf 322} (1989) 235--254.

\bibitem{Barbieri:1995tw}
R.~Barbieri, L.~J. Hall, and A.~Strumia, {\it {Violations of lepton flavor and
  CP in supersymmetric unified theories}},  Nucl. Phys. B {\bf 445} (1995)
  219--251, [\href{http://arxiv.org/abs/hep-ph/9501334}{{\tt hep-ph/9501334}}].

\bibitem{Hisano:1995cp}
J.~Hisano, T.~Moroi, K.~Tobe, and M.~Yamaguchi, {\it {Lepton flavor violation
  via right-handed neutrino Yukawa couplings in supersymmetric standard
  model}},  Phys. Rev. D {\bf 53} (1996) 2442--2459,
  [\href{http://arxiv.org/abs/hep-ph/9510309}{{\tt hep-ph/9510309}}].

\bibitem{Hisano:1998fj}
J.~Hisano and D.~Nomura, {\it {Solar and atmospheric neutrino oscillations and
  lepton flavor violation in supersymmetric models with the right-handed
  neutrinos}},  Phys. Rev. D {\bf 59} (1999) 116005,
  [\href{http://arxiv.org/abs/hep-ph/9810479}{{\tt hep-ph/9810479}}].

\bibitem{Casas:2001sr}
J.~A. Casas and A.~Ibarra, {\it {Oscillating neutrinos and $\mu \to e,
  \gamma$}},  Nucl. Phys. B {\bf 618} (2001) 171--204,
  [\href{http://arxiv.org/abs/hep-ph/0103065}{{\tt hep-ph/0103065}}].

\bibitem{Deppisch:2004fa}
F.~Deppisch and J.~W.~F. Valle, {\it {Enhanced lepton flavor violation in the
  supersymmetric inverse seesaw model}},  Phys. Rev. D {\bf 72} (2005) 036001,
  [\href{http://arxiv.org/abs/hep-ph/0406040}{{\tt hep-ph/0406040}}].

\bibitem{Arganda:2007jw}
E.~Arganda, M.~J. Herrero, and A.~M. Teixeira, {\it {mu-e conversion in nuclei
  within the CMSSM seesaw: Universality versus non-universality}},  JHEP {\bf
  10} (2007) 104, [\href{http://arxiv.org/abs/0707.2955}{{\tt
  arXiv:0707.2955}}].

\bibitem{Altarelli:2012bn}
G.~Altarelli, F.~Feruglio, L.~Merlo, and E.~Stamou, {\it {Discrete Flavour
  Groups, $theta_{13}$ and Lepton Flavour Violation}},  JHEP {\bf 08} (2012)
  021, [\href{http://arxiv.org/abs/1205.4670}{{\tt arXiv:1205.4670}}].

\bibitem{Bambhaniya:2015ipg}
G.~Bambhaniya, P.~S.~B. Dev, S.~Goswami, and M.~Mitra, {\it {The Scalar Triplet
  Contribution to Lepton Flavour Violation and Neutrinoless Double Beta Decay
  in Left-Right Symmetric Model}},  JHEP {\bf 04} (2016) 046,
  [\href{http://arxiv.org/abs/1512.00440}{{\tt arXiv:1512.00440}}].

\bibitem{MEGII:2018kmf}
{\bf MEG II} Collaboration, A.~M. Baldini {\em et.~al.}, {\it {The design of
  the MEG II experiment}},  Eur. Phys. J. C {\bf 78} (2018), no.~5 380,
  [\href{http://arxiv.org/abs/1801.04688}{{\tt arXiv:1801.04688}}].

\bibitem{Bolton:1988af}
R.~D. Bolton {\em et.~al.}, {\it {Search for Rare Muon Decays with the Crystal
  Box Detector}},  Phys. Rev. D {\bf 38} (1988) 2077.

\bibitem{SINDRUM:1987nra}
{\bf SINDRUM} Collaboration, U.~Bellgardt {\em et.~al.}, {\it {Search for the
  Decay mu+ ---\ensuremath{>} e+ e+ e-}},  Nucl. Phys. B {\bf 299} (1988) 1--6.

\bibitem{Mu3e:2020gyw}
{\bf Mu3e} Collaboration, K.~Arndt {\em et.~al.}, {\it {Technical design of the
  phase I Mu3e experiment}},  Nucl. Instrum. Meth. A {\bf 1014} (2021) 165679,
  [\href{http://arxiv.org/abs/2009.11690}{{\tt arXiv:2009.11690}}].

\bibitem{Badertscher:1980bt}
A.~Badertscher {\em et.~al.}, {\it {New Upper Limits for Muon - Electron
  Conversion in Sulfur}},  Lett. Nuovo Cim. {\bf 28} (1980) 401--408.

\bibitem{Mu2e:2014fns}
{\bf Mu2e} Collaboration, L.~Bartoszek {\em et.~al.}, {\it {Mu2e Technical
  Design Report}},  \href{http://arxiv.org/abs/1501.05241}{{\tt
  arXiv:1501.05241}}.

\bibitem{COMET:2018auw}
{\bf COMET} Collaboration, R.~Abramishvili {\em et.~al.}, {\it {COMET Phase-I
  Technical Design Report}},  PTEP {\bf 2020} (2020), no.~3 033C01,
  [\href{http://arxiv.org/abs/1812.09018}{{\tt arXiv:1812.09018}}].

\bibitem{SINDRUMII:1993gxf}
{\bf SINDRUM II} Collaboration, C.~Dohmen {\em et.~al.}, {\it {Test of lepton
  flavor conservation in mu ---\ensuremath{>} e conversion on titanium}},
  Phys. Lett. B {\bf 317} (1993) 631--636.

\bibitem{PRISM:2006abc}
Y.~Kuno {\em et.~al.}, {\it {An Experimental Search for a $\mu N \rightarrow e
  N$ Conversion at Sensitivity of the Order of $10^{−18}$ with a Highly
  Intense Muon Source: PRISM}},  unpublished (2006).

\bibitem{SINDRUMII:2006dvw}
{\bf SINDRUM II} Collaboration, W.~H. Bertl {\em et.~al.}, {\it {A Search for
  muon to electron conversion in muonic gold}},  Eur. Phys. J. C {\bf 47}
  (2006) 337--346.

\bibitem{SINDRUMII:1996fti}
{\bf SINDRUM II} Collaboration, W.~Honecker {\em et.~al.}, {\it {Improved limit
  on the branching ratio of mu ---\ensuremath{>} e conversion on lead}},  Phys.
  Rev. Lett. {\bf 76} (1996) 200--203.

\bibitem{Davidson:2020hkf}
S.~Davidson, {\it {Completeness and complementarity for $\mu \to e\gamma \mu
  \to e \bar e e$ and $\mu A \to eA$}},  JHEP {\bf 02} (2021) 172,
  [\href{http://arxiv.org/abs/2010.00317}{{\tt arXiv:2010.00317}}].

\bibitem{Cirigliano:2017azj}
V.~Cirigliano, S.~Davidson, and Y.~Kuno, {\it {Spin-dependent $\mu \to e$
  conversion}},  Phys. Lett. B {\bf 771} (2017) 242--246,
  [\href{http://arxiv.org/abs/1703.02057}{{\tt arXiv:1703.02057}}].

\bibitem{Davidson:2018kud}
S.~Davidson, Y.~Kuno, and M.~Yamanaka, {\it {Selecting $\mu \to e$ conversion
  targets to distinguish lepton flavour-changing operators}},  Phys. Lett. B
  {\bf 790} (2019) 380--388, [\href{http://arxiv.org/abs/1810.01884}{{\tt
  arXiv:1810.01884}}].

\bibitem{Okada:1999zk}
Y.~Okada, K.-i. Okumura, and Y.~Shimizu, {\it {Mu --\ensuremath{>} e gamma and
  mu --\ensuremath{>} 3 e processes with polarized muons and supersymmetric
  grand unified theories}},  Phys. Rev. D {\bf 61} (2000) 094001,
  [\href{http://arxiv.org/abs/hep-ph/9906446}{{\tt hep-ph/9906446}}].

\bibitem{Aiba:2021bxe}
M.~Aiba {\em et.~al.}, {\it {Science Case for the new High-Intensity Muon Beams
  HIMB at PSI}},  \href{http://arxiv.org/abs/2111.05788}{{\tt
  arXiv:2111.05788}}.

\bibitem{Kuno:1999jp}
Y.~Kuno and Y.~Okada, {\it {Muon decay and physics beyond the standard model}},
   Rev. Mod. Phys. {\bf 73} (2001) 151--202,
  [\href{http://arxiv.org/abs/hep-ph/9909265}{{\tt hep-ph/9909265}}].

\bibitem{Okada:1997fz}
Y.~Okada, K.-i. Okumura, and Y.~Shimizu, {\it {CP violation in the mu
  ---\ensuremath{>} 3 e process and supersymmetric grand unified theory}},
  Phys. Rev. D {\bf 58} (1998) 051901,
  [\href{http://arxiv.org/abs/hep-ph/9708446}{{\tt hep-ph/9708446}}].

\bibitem{Ilakovac:2012sh}
A.~Ilakovac, A.~Pilaftsis, and L.~Popov, {\it {Charged lepton flavor violation
  in supersymmetric low-scale seesaw models}},  Phys. Rev. D {\bf 87} (2013),
  no.~5 053014, [\href{http://arxiv.org/abs/1212.5939}{{\tt arXiv:1212.5939}}].

\bibitem{Koike:2010xr}
M.~Koike, Y.~Kuno, J.~Sato, and M.~Yamanaka, {\it {A new idea to search for
  charged lepton flavor violation using a muonic atom}},  Phys. Rev. Lett. {\bf
  105} (2010) 121601, [\href{http://arxiv.org/abs/1003.1578}{{\tt
  arXiv:1003.1578}}].

\bibitem{Uesaka:2016vfy}
Y.~Uesaka, Y.~Kuno, J.~Sato, T.~Sato, and M.~Yamanaka, {\it {Improved analyses
  for $\mu^-e^-\rightarrow e^-e^-$ in muonic atoms by contact interactions}},
  Phys. Rev. D {\bf 93} (2016), no.~7 076006,
  [\href{http://arxiv.org/abs/1603.01522}{{\tt arXiv:1603.01522}}].

\bibitem{Uesaka:2017yin}
Y.~Uesaka, Y.~Kuno, J.~Sato, T.~Sato, and M.~Yamanaka, {\it {Improved analysis
  for $\mu^-e^-\to e^-e^-$ in muonic atoms by photonic interaction}},  Phys.
  Rev. D {\bf 97} (2018), no.~1 015017,
  [\href{http://arxiv.org/abs/1711.08979}{{\tt arXiv:1711.08979}}].

\bibitem{Kuno:2019ttl}
Y.~Kuno, J.~Sato, T.~Sato, Y.~Uesaka, and M.~Yamanaka, {\it {Momentum
  distribution of the electron pair from the charged lepton flavor violating
  process $\mu^-e^-\to e^-e^-$ in muonic atoms with a polarized muon}},  Phys.
  Rev. D {\bf 100} (2019), no.~7 075012,
  [\href{http://arxiv.org/abs/1908.11653}{{\tt arXiv:1908.11653}}].

\end{thebibliography}\endgroup


\providecommand{\href}[2]{#2}\begingroup\raggedright\endgroup

\end{document}